%% file: ANA-STDM-2023-01-PAPER.tex
\documentclass[PAPER, atlasdraft=false, cernpreprint, UKenglish, texmf, orcidlogo]{atlasdoc}
 
\usepackage[subfigure]{atlaspackage}
\usepackage{atlasbiblatex}
 
\usepackage{atlasphysics}

\graphicspath{{logos/}{figures/}}
 
\usepackage{ANA-STDM-2023-01-PAPER-defs}

\AtlasTitle{A precise determination of the strong-coupling constant from the recoil of $Z$ bosons with the ATLAS experiment at $\sqrt{s} = 8$~TeV}

\AtlasAbstract{
The coupling constant of the strong force is determined from
the transverse-momentum distribution of $Z$ bosons produced in 8~TeV
proton--proton collisions at the LHC and recorded by the ATLAS experiment.
The $Z$-boson cross sections are measured in the full phase space of
the decay leptons using 15.3~million electron and muon pairs, in a
dataset collected in 2012 and corresponding to an integrated
luminosity of 20.2~fb$^{-1}$.
The analysis is
based on predictions evaluated at third order in perturbative QCD,
supplemented by the resummation of logarithmically enhanced
contributions in the low transverse-momentum region of the lepton pairs.
The determined value of the strong coupling at the reference scale
corresponding to the $Z$-boson mass is $\alpha_\text{s}(m_Z) =
0.1183 \pm 0.0009$.
This is the most precise experimental determination of
$\alpha_\text{s}(m_Z)$ achieved so far.
}

\AtlasRefCode{STDM-2023-01}
 
\PreprintIdNumber{CERN-EP-2023-200}

\AtlasJournal{Nature Phys.}

\hypersetup{pdftitle={ATLAS document},pdfauthor={The ATLAS Collaboration}}

\begin{document}
 
\maketitle

\section{Introduction}
The coupling constant of the strong interaction is one of the fundamental
parameters of the Standard Model, yet remains the least precisely
determined of the four fundamental couplings in nature.
The strong interaction is described theoretically by quantum
chromodynamics (QCD), a gauge field theory with symmetry group
SU(3)~\cite{Gross:1973id,Politzer:1973fx}. 
The free parameters of QCD include the six quark masses and the
running coupling constant $\alphas(Q)$, which runs with the energy
scale $Q$ characterising the interaction.
While the running of the coupling constant is fully predicted by theory,
its value at a reference scale needs to be determined experimentally.
The latest world-average value of experimental determinations and
QCD lattice calculations of the strong-coupling constant at the
scale of the $Z$-boson mass is calculated to be $\alphas(Q=\mZ) = 0.1179 \pm 0.0009$,
with a relative uncertainty of
$0.8\%$~\cite{ParticleDataGroup:2022pth}.
This uncertainty is orders of magnitude larger than that of the
couplings of the other three fundamental interactions: the
electromagnetic, weak, and gravitational forces.
 
Our knowledge of the strong-coupling constant has improved over the years,
from the significant uncertainties in the first determinations in the
mid 1980s~\cite{ParticleDataGroup:1986kuw}, to the present uncertainty at the percent level.
Further improvement in the precision of $\alphas$ is important in order
to reduce the associated theoretical uncertainty which enters
all cross-section calculations for processes at the LHC and affects
several key observables at $\epem$ colliders.
As an example, in the global fit of the electroweak sector of the
Standard Model, the value of \asmz{} is the leading source of
uncertainty in the computation of the total and partial hadronic decay widths
of the $Z$ boson~\cite{Heinemeyer:2021rgq,Haller:2018nnx,dEnterria:2020cpv}.
A precise determination of \asmz{} is required to fully exploit the
sensitivity to new physics expected from high-precision measurements of such
observables at future $\epem$ colliders.
The value of $\asmz$ and its energy evolution also have far-reaching
implications for the stability of the electroweak
vacuum~\cite{Degrassi:2012ry} 
and the convergence of the couplings of the strong, weak and electromagnetic forces
at an energy close to the Planck scale, which might signal the onset
of a grand unification of these forces.
 
Various different determinations of $\asmz$ contribute to the current world average, and
are categorised according to their methodological approach~\cite{doi:10.1142/9789813238053_0007}. The most
precise determinations are based on lattice QCD analysis of
hadron spectroscopy, resulting in $\asmz = 0.1184 \pm
0.0008$~\cite{Aoki:2021kgd}, and hadronic $\tau$-lepton decays, resulting in $\asmz =
0.1177 \pm 0.0019$~\cite{Baikov:2008jh,Pich:2016bdg,Davier:2013sfa,Boito:2018yvl,Boito:2020xli,ParticleDataGroup:2020ssz,ParticleDataGroup:2022pth}.
Arguably the cleaner determinations are those from global fits of the electroweak
observables, which exploit the sensitivity of total and partial hadronic decay widths
of the $Z$ boson, as in Ref.~\cite{Haller:2018nnx}, yielding $\asmz = 0.1194 \pm
0.0029$. These determinations have been performed at
next-to-next-to-next-to-leading order (N$^3$LO) in QCD and are
currently limited by experimental uncertainties.
In hadron-induced collisions, the strong-coupling constant has been determined
from final states with
jets~\cite{CMS-SMP-20-011,STDM-2018-51,H1:2017bml,Britzger:2019kkb},  
from inclusive top-quark pair
production~\cite{CMS-TOP-17-001,Klijnsma:2017eqp}, and 
from inclusive $W$- and $Z$-boson
production~\cite{dEnterria:2019aat}. 
The high-momentum region of the $Z$-boson transverse-momentum (\pt)
distribution measured at the
LHC~\cite{STDM-2012-23,STDM-2014-12,CMS-SMP-13-013} was
included in the determination of parton distribution functions (PDFs)~\cite{Boughezal:2017nla}, and contributed to the
simultaneous determination of PDFs and the strong-coupling constant in Refs.~\cite{Ball:2018iqk,Hou:2019efy,Cridge:2021qfd}.
 
Further improvement in our knowledge of \asmz{} is limited by two important
sources of theoretical uncertainty: the accuracy of the perturbative
predictions and the size of non-perturbative effects.
In this context, it is highly desirable to investigate alternative
determinations of \asmz{} based on the most sensitive observables and
state-of-the-art theory predictions.
 
This article presents a precise determination of $\asmz$ from a
semi-inclusive observable,\footnote{Semi-inclusive observables are
those with more than one kinematic momentum scale in the
perturbative regime, where the semi-inclusive region is close to the boundary of the
phase space allowed by the kinematics. In such a limit, the associated
parton radiation is strongly inhibited and large logarithmic
corrections appear in the perturbative
computation~\cite{Catani:1989ne,Catani:1990rr}.} namely the
low-momentum Sudakov region\footnote{The low-energy region of the
transverse-momentum distribution of $Z$ bosons is characterised by
very high probabilities of gluon emissions with vanishingly small
momenta. Rather than calculate each of these, it is theoretically
simpler to model them as a single factor quantifying the probability
of no emission, known as the Sudakov form factor~\cite{Sudakov:1954sw}.} of the
transverse-momentum distribution of $Z$ bosons produced via the
Drell--Yan process~\cite{Drell:1970wh}, which refers to the production of
a massive lepton-pair in hadron--hadron collisions at high energies.
The strong force is responsible for the radiation from the
initial-state partons, and for the subsequent recoil of the $Z$ bosons
which acquire non-zero transverse momentum with respect to the incoming proton beam axis.
The hardness of the transverse-momentum distribution is a measure of
the strength of the recoil of the $Z$ bosons, which in turn is
proportional to the strong coupling.
In contrast to most other determinations of \asmz{} at hadron colliders,
which analyse observables based on QCD final-state radiative objects, this analysis uses QCD initial-state radiative processes
to determine the strong coupling.
In the Drell--Yan process, the final-state particles are not subject
to the strong interaction, which reduces theoretical complications and
uncertainties.
The energy scale at which
the strong-coupling constant is perturbatively expanded is unambiguously fixed to the $Z$-boson mass.
This methodology was tested in Ref.~\cite{Camarda:2022qdg}
using proton--antiproton collision data at the Tevatron, and is
applied here for the first time at the LHC.
 
\begin{figure}
\centering
\includegraphics[width=0.5\textwidth]{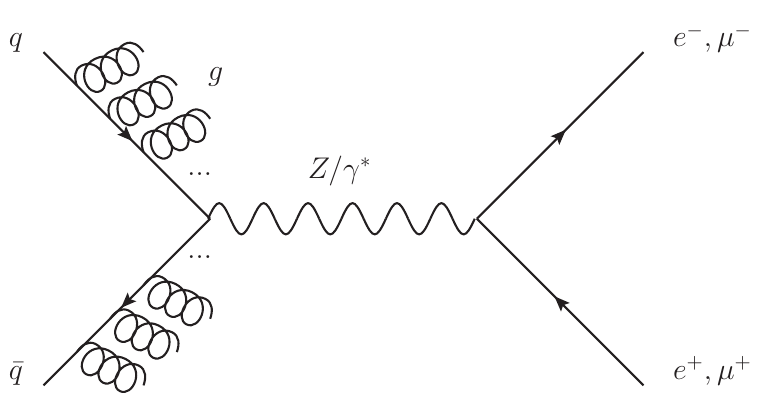}
\caption{Leading-order Feynman diagram for the production of a massive
electron or muon pair through the Drell--Yan process, including soft gluon
radiation from the initial-state quarks.\label{fig:feynman}}
\end{figure}

\begin{figure}
\centering
\includegraphics[width=0.7\textwidth]{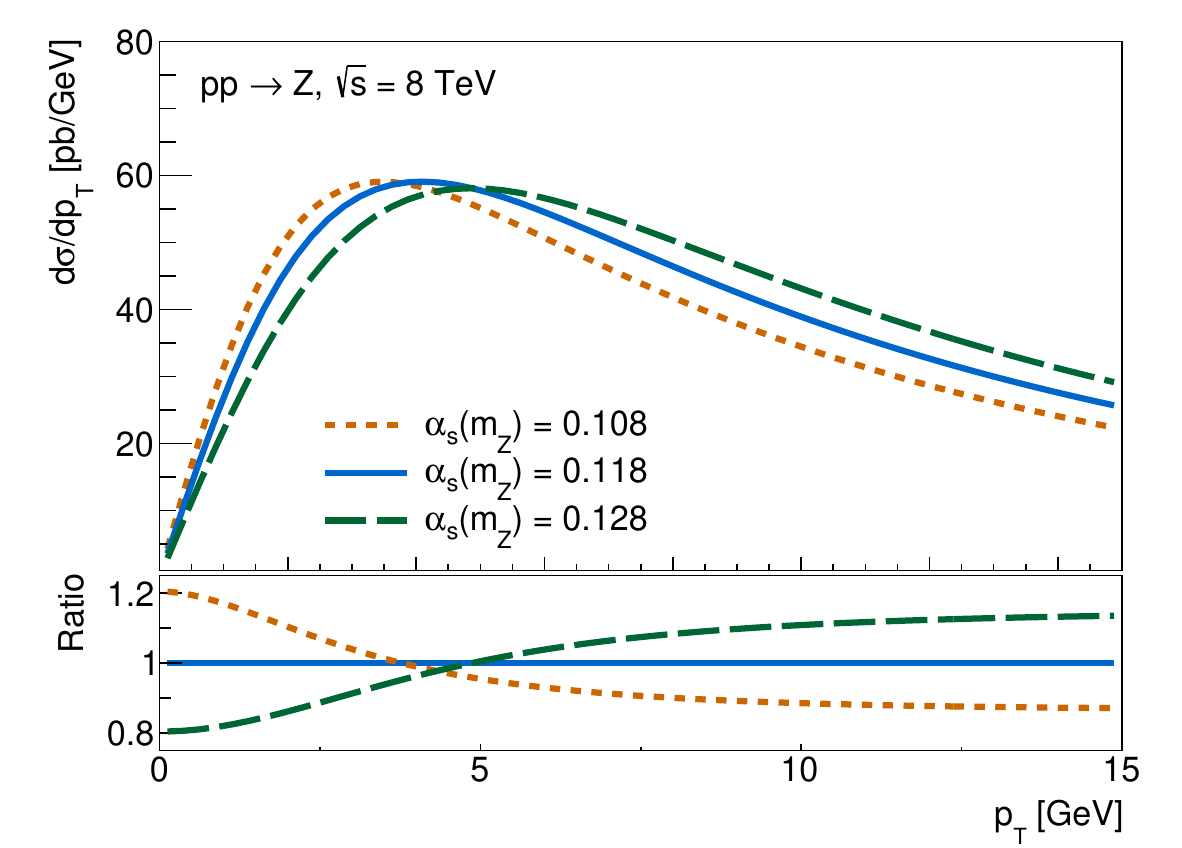}
\caption{Transverse-momentum distribution of $Z$ bosons predicted
with \texttt{DYTurbo}~\cite{Camarda:2019zyx} at different values of $\alpha_\text{s}(m_Z)$, using the MSHT20
PDF set~\cite{Bailey:2020ooq}. The impact of changing $\alpha_\text{s}(m_Z)$ on the PDFs is included.\label{fig:sensitivity}}
\end{figure}

Figure~\ref{fig:feynman} depicts the leading-order Feynman diagram of
the Drell--Yan process, with a schematic representation of soft gluon radiation from the
initial-state quarks. Figure~\ref{fig:sensitivity} shows the $Z$-boson
transverse-momentum distribution for three different values of \asmz.
 
Compared to other determinations of $\asmz$
at hadron colliders, based on either exclusive or
inclusive observables, this determination gathers the desirable
features for high precision: large observable sensitivity to
$\asmz$ relative to the experimental precision, and high
perturbative accuracy of the theoretical
predictions~\cite{Camarda:2021ict,Chen:2022cgv,Neumann:2022lft},
enabled by the computation of some perturbative corrections in QCD at
four- or five-loop level~\cite{Baikov:2016tgj,Herzog:2018kwj,Duhr:2022yyp,Moult:2022xzt,Chakraborty:2022yan}.

\section{ATLAS detector and data sample}
 
\newcommand{\AtlasCoordFootnote}{
ATLAS uses a right-handed coordinate system with its origin at the nominal interaction point (IP)
in the centre of the detector and the \(z\)-axis along the beam pipe.
The \(x\)-axis points from the IP to the centre of the LHC ring,
and the \(y\)-axis points upwards.
Cylindrical coordinates \((r,\phi)\) are used in the transverse plane,
\(\phi\) being the azimuthal angle around the \(z\)-axis.
The pseudorapidity is defined in terms of the polar angle \(\theta\) as \(\eta = -\ln \tan(\theta/2)\).
Angular distance is measured in units of \(\Delta R \equiv \sqrt{(\Delta\eta)^{2} + (\Delta\phi)^{2}}\).}

The ATLAS experiment~\cite{PERF-2007-01} at the LHC is a multipurpose particle detector
with a forward--backward symmetric cylindrical geometry and a near \(4\pi\) coverage in
solid angle.\footnote{\AtlasCoordFootnote}
It consists of an inner tracking detector surrounded by a thin superconducting solenoid
providing a \qty{2}{\tesla} axial magnetic field, electromagnetic and hadron calorimeters, and a muon spectrometer.
The inner tracking detector covers the pseudorapidity range \(|\eta| < 2.5\).
It consists of silicon pixel, silicon microstrip, and transition radiation tracking detectors.
Lead/liquid-argon (LAr) sampling calorimeters provide electromagnetic (EM) energy measurements
with high granularity.
A steel/scintillator-tile hadron calorimeter covers the central pseudorapidity range (\(|\eta| < 1.7\)).
The endcap and forward regions are instrumented with LAr calorimeters
for both the EM and hadronic energy measurements up to \(|\eta| = 4.9\).
The muon spectrometer surrounds the calorimeters and is based on
three large superconducting air-core toroidal magnets with eight coils each.
The field integral of the toroids ranges between \num{2.0} and \qty{6.0}{\tesla\metre}
across most of the detector.
The muon spectrometer includes a system of precision chambers for tracking and fast detectors for triggering.
A three-level trigger system is used to select events.
The first-level trigger is implemented in hardware and uses a subset of the detector information
to accept events at a rate of at most \qty{75}{\kilo\hertz}.
This is followed by two software-based trigger levels that
together reduce the accepted event rate to \qty{400}{\hertz} on average
depending on the data-taking conditions during 2012.
An extensive software suite~\cite{ATL-SOFT-PUB-2021-001} is used in data simulation, in the reconstruction
and analysis of real and simulated data, in detector operations, and in the trigger and data acquisition
systems of the experiment.
The data were collected by the ATLAS detector in 2012 at a
centre-of-mass energy of $\sqrt{s} = 8$~TeV, and correspond to an
integrated luminosity of~20.2~fb$^{-1}$.
The mean number of additional
$pp$ interactions per bunch crossing (pile-up events) in the dataset
is approximately~20.
 
\section{Cross-section measurement}
 
The $Z$-boson transverse-momentum distribution is measured
in the electron and muon decay channels, which provide a clear
signature with low background rates and a high-precision measurement of the
momentum, as presented in Ref.~\cite{STDM-2018-05}.
The double-differential cross sections as functions of
transverse momentum (\pt) and rapidity ($y$) of the \Zboson{} boson are measured in the pole
region, defined as~$80 < m_{\ell\ell} < 100$~\GeV, where $m_{\ell\ell}$ is the
invariant mass of the dilepton system.
The combination of 6.2~million electron pairs and 7.8~million muon pairs
in the central region, with pseudorapidity $\abseta < 2.4$, is
complemented by 1.3~million electron pairs with one electron in the
forward region of the detector, $2.5 < \abseta < 4.9$, and one
electron in the central region.
 
Events with two \enquote{central} electrons were collected using either a
dielectron trigger, requiring two electron candidates to each have
$\pT > 12$~\GeV{}, or two high-\pT{} single-electron triggers, the
main one having a \pT{} threshold of 24~\GeV.
Events with two central muons were collected with either a dimuon
trigger requiring two muon candidates with $\pT > 18$~\GeV{} and
8~\GeV, respectively, or two high-\pT{} single-muon triggers, the
main one having a \pT{} threshold of 24~\GeV. Events with
one central electron and one forward electron were collected using the two
high-\pT{} single-electron triggers described above.
Electron candidates in the central region are required in the offline analysis to
have either $\pT > 20$~\GeV{} when paired with another central electron
candidate or $\pT > 25$~\GeV{} when paired with a forward electron
candidate.
Muon candidates are required offline to have $\pT > 20$~\GeV.
Electron candidates in the forward region are required offline to have
$\pT > 20$~\GeV.
 
The cross-section measurement relies on the decomposition of the leptons' \costheta\ and
$\phi$ angular distributions in the Collins--Soper frame\footnote{The
Collins--Soper frame is the rest frame of the \Zboson boson in
which the $z$-axis bisects the angle between the momentum vector of one
proton and the negative momentum vector of the other proton, the $x$-axis is
fixed by the transverse momentum of the \Zboson boson, and the $y$-axis
completes a right-handed coordinate system~\cite{Collins:1977iv}.}
into nine spherical harmonic polynomials, multiplied by angular
coefficients~\cite{Mirkes:1992hu}. 
The cross sections in the full lepton phase space, i.e.\ without
kinematic requirements on the transverse momenta and pseudorapidity of
the leptons, are extracted from the data by fitting templates of the
spherical harmonic polynomials to the reconstructed angular
distributions in \costhetaphi.
The decomposition into nine spherical harmonics is based on a simple
and model-independent ansatz, namely the spin-one nature of the intermediate
boson and spin-half nature of the decay leptons, and on the assumption
of angular momentum conservation and quantisation. By measuring the cross sections in the
full lepton phase space there is no longer a requirement for
predictions to model the polarisation and decay of the \Zboson{}
boson; only its production properties are of interest for comparison with the
measurements, thus enabling a determination of the strong-coupling
constant not affected by theoretical uncertainties and ambiguities
related to spin correlations.
 
The double-differential cross sections are measured in eight rapidity
regions in the range $|y| < 3.6$. The region of $Z$-boson transverse
momentum $\pt < 29$~\GeV{} is considered for the determination
of \asmz, corresponding to nine bins of transverse momentum. The
choice of bin boundaries in \pt{} is the result of an optimisation
with respect to the limited resolution of the measurements at low \pt.
The background rates from other QCD and electroweak processes are at
the level of $0.3\%$ for central electron and muon pairs, and $1.1\%$
for events with one central electron and one forward electron.
The total uncertainties in the measurements are dominated by the
statistical uncertainties of the data and, to a lesser extent, those of the
simulation samples. Apart from the luminosity uncertainty of $1.8\%$~\cite{DAPR-2013-01},
the total uncertainties are below $1\%$ for $|y| < 2.0$, and
below $10\%$ for $2.0 < |y| < 3.6$~\cite{STDM-2018-05}.

\section{Theoretical framework and statistical analysis}
 
The theoretical predictions are computed with the public numerical program
\texttt{DYTurbo}~\cite{Camarda:2019zyx}, which implements the resummation of logarithmically enhanced
contributions in the low-\qt{} region of the lepton pairs at
approximate next-to-next-to-next-to-next-to-leading-logarithm accuracy (N$^4$LLa)~\cite{n4lla},
combined with the hard-collinear contributions at
N$^3$LO in powers of the QCD
coupling~\cite{Camarda:2021ict}, and matched to fixed order, namely to N$^3$LO.
The resummation is carried out in impact-parameter space $b$, which is the
Fourier-transform variable conjugate to \qt~\cite{Collins:1984kg,Bozzi:2005wk,Catani:2015vma}.
The resummed cross section is given by the convolution of the
leading-order (LO) cross section, the hard-collinear contributions,
and the universal (process-independent) Sudakov form factor.
The hard-collinear contributions are expanded in powers of \as{},
whereas the Sudakov form factor contains all the terms that
order-by-order in $\as$ are logarithmically divergent as $\qt \to 0$.
A unitarity constraint is imposed in the matching to
fixed order of the \qt-resummed prediction so as to exactly recover the
N$^3$LO finite-order result when integrating the full lepton-phase-space resummed cross section over \qt.
The $\mathcal{O}(\as^3)$ coefficient of the $Z$+jet cross-section
predictions, required for the matching to fixed order, was computed with
MCFM~\cite{Boughezal:2015ded,Neumann:2022lft}, using a lower cut-off of
$\pt = 5$~\GeV. The corresponding matching corrections were
interpolated with their known quadratic dependence on the ratio of
$\pt/m_Z$~\cite{Camarda:2021jsw} and extrapolated to $\pt = 0$.
The Sudakov form factor is singular in the region of transverse momenta
of the order of the QCD coupling scale
$\Lambda_{\textrm{QCD}}$. This signals that a truly non-perturbative
region is approached and perturbative results are not reliable.
Non-perturbative QCD effects are included with a corresponding form
factor~\cite{Collins:1984kg,Collins:2014jpa}, which depends on a set
of parameters which are either left free in the fit for the
determination of $\asmz$ or varied when
assessing non-perturbative uncertainties.
 
The predicted cross sections depend on three unphysical QCD scales: the
renormalisation scale \muR, which refers to the characteristic
energy scale at which the running coupling constant is evaluated, the
factorisation scale \muF, which separates long-distance and short-distance physics in a scattering
process, and the resummation scale $Q$, which parameterises the
arbitrariness in the resummation procedure. The central value of each scale is set to
the quadratic sum of
$m_{\ell\ell}$ and \pT of the $Z$ boson.
 
The PDF set used in the predictions is the approximate N$^3$LO MSHT20
PDF set~\cite{McGowan:2022nag}, which is the only PDF set currently
available at this order.
The PDFs are interpolated with LHAPDF~\cite{Buckley:2014ana} at the factorisation
scale \muF, and evolved backwards using the N$^3$LO solution of the
evolution equation. 
The number of active flavours is set to five in all the coefficients
entering the calculation, and in the evolution of the PDFs. The charm-
and bottom-quark PDFs are asymptotically switched off in the backward
evolution when approaching their corresponding thresholds.
 
The effect of initial-state radiation of photons on the
transverse-momentum distribution's shape is estimated at leading-logarithm accuracy
with \PYTHIA[8]~\cite{Sjostrand:2014zea} and the AZ set of tuned
parton shower parameters~\cite{STDM-2012-23}, and applied as a
bin-by-bin multiplicative correction factor. A computation of initial-state radiation
of photons at next-to-leading-logarithm
accuracy~\cite{Cieri:2018sfk} is used to validate 
the \PYTHIA[8] predictions.
Higher-order effects on the cross-section normalisation from QED
initial-state radiation and from electroweak virtual corrections are
considered at next-to-leading order. These are
directly computed using the code from~Ref.~\cite{Bondarenko:2022mbi},
and are in agreement with the results from other calculations
benchmarked by the LHC EW working group. At the $Z$~pole, the virtual
effects decrease the predicted cross sections by~0.8\%, while the
QED initial-state effects increase them by~0.4\%. These corrections
are found to be independent of rapidity.
Higher-order electroweak corrections are expected to be very
small at the \Zboson-boson pole, and are neglected.\footnote{The electroweak parameters are set according to the $G_\mu$ scheme,
in which the Fermi coupling constant $G_\textrm{F}$, the $W$-boson mass $m_W$,
and the $Z$-boson mass $m_Z$ are set to the input values
$G_\textrm{F} = 1.1663787 \cdot 10^{-5}$~$\GeV^{-2}$, $m_W =
80.385$~\GeV, and $m_Z =
91.1876$~\GeV~\cite{ParticleDataGroup:2020ssz}, whereas the weak-mixing
angle and the QED coupling are calculated at tree level.}
 
The statistical analysis for the determination of $\asmz$ is
performed in the xFitter framework~\cite{Alekhin:2014irh}.
The value of $\asmz$ is determined by minimising a $\chi^2$
function which includes both the experimental uncertainties and the
theoretical uncertainties arising from PDF variations:
\begin{eqnarray}
\nonumber \lefteqn{\chi^2(\beta_{\textrm{exp}},\beta_{\textrm{th}}) = }  \\
&& \nonumber \sum_{i=1}^{N_{\textrm{data}}} \frac{\textstyle \left( \sigma^{\textrm{exp}}_i + \sum_j \Gamma^{\textrm{exp}}_{ij} \beta_{j,\textrm{exp}} - \sigma^{\textrm{th}}_i - \sum_k \Gamma^{\textrm{th}}_{ik}\beta_{k,\textrm{th}} \right)^2}{\Delta_i^2} \\
&&   + \sum_j \beta_{j,\textrm{exp}}^2 + \sum_k \beta_{k,\textrm{th}}^2\,.   \label{eq:chi2prof}
\end{eqnarray}
The correlated experimental and theoretical uncertainties are included
by using the nuisance parameter vectors $\beta_{\textrm{exp}}$
and $\beta_{\textrm{th}}$, respectively. Their influence on
the data and theory predictions is described by the $\Gamma^{\textrm{exp}}_{ij}$ and $\Gamma^{\textrm{th}}_{ik}$ matrices. The index $i$ runs
over all $N_{\textrm{data}} = 72$ data points of the double-differential
\Zboson-boson \pT{} and $y$ distribution, whereas the indices $j$ and
$k$ correspond to the experimental and theoretical uncertainty
nuisance parameters respectively.
The measurements and the uncorrelated experimental
uncertainties are given by $\sigma^{\textrm{exp}}_i$ and $\Delta_i$\,, respectively, and
the theory predictions are $\sigma_i^{\textrm{th}}$.
The matrices $\Gamma^{\textrm{exp}}_{ij}$ encode all the information
in the experimental covariance matrix of the measured
double-differential cross sections as functions of the transverse momentum and
rapidity of the \Zboson{} boson. The matrices
$\Gamma^{\textrm{th}}_{ik}$ cover the nuisance parameters of the PDF
Hessian uncertainties, and parameters of the non-perturbative form
factor, which are left free in the fit by allowing unconstrained variations.
 
Determinations of \asmz{} at hadron colliders are usually affected by
significant correlations between \asmz{} and the PDFs, especially
the gluon PDF~\cite{H1:2015ubc}. The dependence of the PDFs on the value of
$\asmz$ is accounted for by using corresponding
$\as$-series of PDF sets, which are provided for seven fixed values
of \asmz{} in the range $0.114 < \asmz < 0.120$.
At each value of $\asmz$, the PDF uncertainties
are Hessian profiled and the $\chi^2$ function is minimised by solving
a system of linear equations, according to
Eq.~(\ref{eq:chi2prof})~\cite{HERAFitterdevelopersTeam:2015cre},
whereas the different values of $\chi^2$ as a function of \asmz{} are minimised through a
polynomial interpolation to determine \asmz.
 
A validation of the statistical analysis, as well as an estimate of the
sensitivity of the measured  \Zboson-boson cross sections to \asmz, is
provided by a pseudo-fit.
Identical theory predictions are used as central values for both data
and theory in Eq.~(\ref{eq:chi2prof}), including all statistical and
systematic experimental uncertainties, without theoretical
uncertainties, and with fixed values of the non-perturbative QCD
parameters.
The input value is set to $\asmz = 0.118$, and the pseudo-fit yields
$\asmz = 0.11801 \pm 0.00006$.
The closure of the method is thus found to be accurate to $0.01\%$ and
the relative uncertainty in \asmz{} is estimated to be $0.05\%$ before
including the theoretical uncertainties discussed in the following.
 
\section{Determination of \asmz}
The determination of $\asmz$ with central values of the QCD scales and
using Eq.~(\ref{eq:chi2prof}) yields $\asmz = 0.11847 \pm 0.00067$,
with contributions to the fit uncertainty from the experimental
sources and from the PDFs estimated as $\pm 0.00044$ and $\pm 0.00051$
respectively.\footnote{The non-perturbative QCD parameters are left free in the
fit, and, due to their correlation with $\asmz$, the experimental
uncertainties are significantly larger here than in the pseudo-fit, where they
are kept fixed to assess the experimental sensitivity to \asmz.}
Uncertainties arising from missing higher orders, due to truncation of
the perturbative series, are estimated through independent variations
of \muR, \muF and $Q$ in the range
$0.5 \cdot \sqrt{m_{\ell\ell}^2+\pT^2} \leq \{ \muR, \muF, Q \} \leq 2  \cdot \sqrt{m_{\ell\ell}^2+\pT^2}$ with
the constraints $0.5\leq \{ \muF/\muR, Q/\muR, Q/\muF \}\leq 2$,
leading to 14 variations.
 
The fit is repeated for each scale variation, and the determined values
of $\asmz$ range from a minimum of $0.11786$ to a maximum of
$0.11870$. The midpoint of this range of $\asmz$ values, $\asmz =
0.11828$, is taken as the nominal result, and the range's envelope
of $\pm 0.00042$ is used as an estimate of the uncertainties due to missing higher orders,
henceforth referred to as \enquote{missing higher-order uncertainties}.
 
\begin{figure}
\centering
\includegraphics[width=0.7\textwidth]{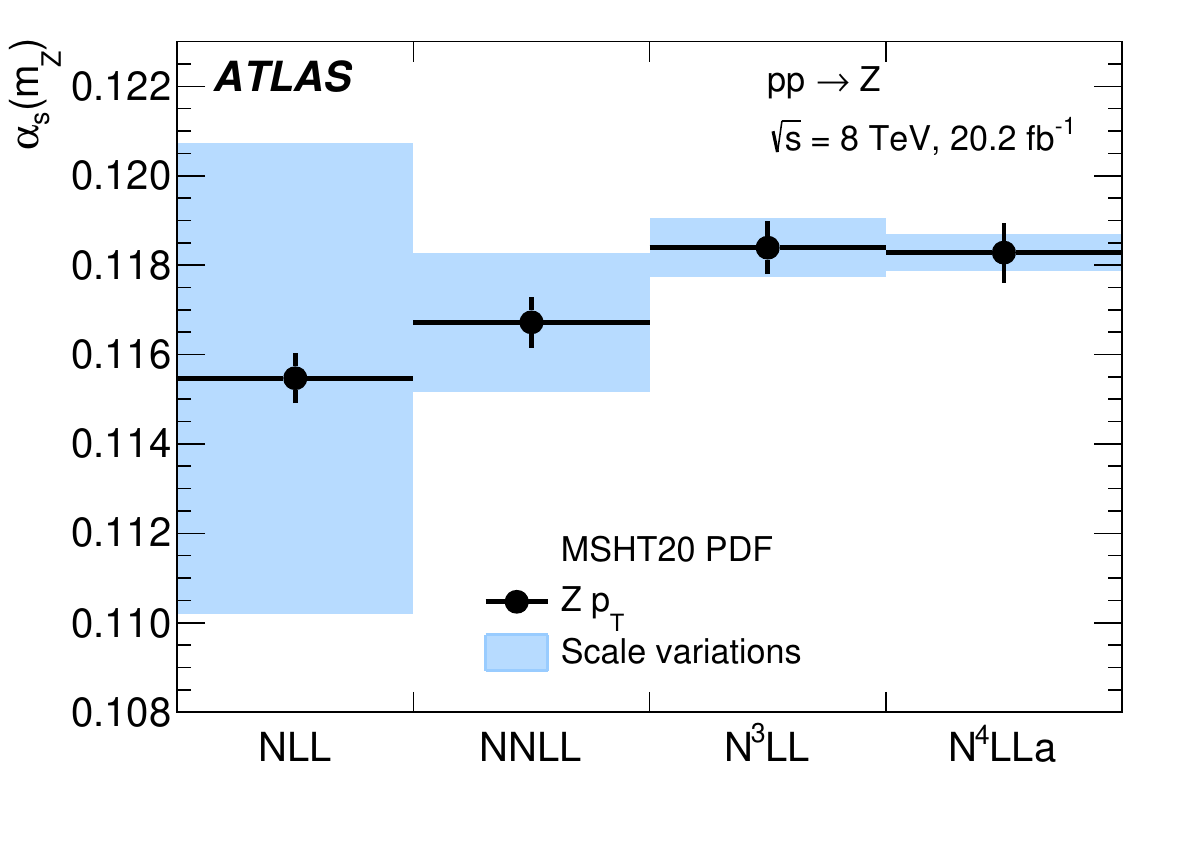}
\caption{Determination of \asmz{} at various different orders in
the QCD perturbative expansion, using the MSHT20 PDF set. The filled
area represents missing higher-order uncertainties estimated through
scale variations, and the vertical error bars include experimental and PDF
uncertainties.\label{fig:orders}}
\end{figure}

The procedure is repeated at lower orders, starting from
next-to-leading-logarithm accuracy matched to next-to-leading order
(NLL+NLO). The MSHT20 PDF set is used throughout, and the order of the
PDFs is matched to the order required by the logarithmic terms included
in the \pt-resummation, i.e.\ NNLO at N$^3$LL and NLO at
NNLL.\footnote{At NLL the NLO PDF set is used because the LO PDF set
does not have \asmz{} variations.} The results are shown in
Figure~\ref{fig:orders}.
At every order, the estimate of missing higher-order uncertainties
obtained from the scale variations overlaps with determinations of \asmz{} at
higher orders, giving confidence in the robustness and gradual
convergence of these estimates.
 
Fits without the $\mathcal{O}(\as^3)$ matching corrections yield an
$\asmz$ central value which is $0.00024$ lower, and
the half envelope due to the scale variations increases from $\pm 0.00042$ to
$\pm 0.00062$, which is consistent with the observed shift.
Uncertainties in the matching to fixed order are estimated
with fits in which the unitarity constraint is not applied.
For these fits, the midpoint and half envelope of $\asmz$ values
from the scale variations yield $\asmz = 0.11820 \pm 0.00037$.
The difference between this set of fits and the nominal set of fits is taken
as a one-sided matching uncertainty of $-0.00008$.
 
Additional uncertainties in the modelling of the non-perturbative form factor
are estimated with variations of corresponding parameters, leading to
an estimate of $^{+0.00012}_{-0.00020}$, as described in Section~\ref{sec:methods}.
The effect of charm- and bottom-quark masses and thresholds are
estimated with various alternative fits, such as by including variable-flavour
number either in the evolution of the PDFs ($-0.00029$) or in the running
of \as~\cite{Herren:2017osy} in the Sudakov form factor ($+0.00021$), by varying
the charm threshold $\mu_c$ by a factor of 2 ($+0.00007$), by varying the bottom
threshold $\mu_b$ by a factor of 0.5 ($-0.00029$), or by including
the effect of final-state gluon-splitting into massive bottom-quark ($+0.00040$) and
charm-quark ($+0.00001$) pairs. The largest excursions are taken as an
estimated uncertainty of $^{+0.00040}_{-0.00029}$ associated with the flavour model.
 
The inclusion of initial-state radiation of photons at leading-logarithm accuracy
shifts the value of \asmz{} by $-0.00028$. Half of this shift
is assigned as an uncertainty associated with missing
higher-order corrections for the initial-state radiation of
photons. Initial-state radiation of photons at next-to-leading-logarithm
accuracy~\cite{Cieri:2018sfk} shifts the value of \asmz{} by
$+0.00007$, which is well within the assigned uncertainty.
The inclusion of NLO electroweak corrections shifts the value of \asmz{}
by $+0.00006$, and uncertainties related to missing electroweak higher
orders are considered negligible.
 
Uncertainties related to the numerical approximation or our incomplete
knowledge of some of the coefficients required for N$^4$LL accuracy of \qt-resummation
are estimated to contribute at the level of $\pm0.00004$, with the
largest contribution coming from the numerical approximation of the
\textit{cusp} anomalous dimension at five loops~\cite{Herzog:2018kwj}, and from
our incomplete knowledge of the hard-collinear contributions at four
loops~\cite{Chakraborty:2022yan}. Uncertainties due to the numerical approximation of the four-loop
splitting functions are already included in the MSHT20 PDF
uncertainties. 
 
A summary of the uncertainties in the determination of \asmz{} is shown
in Table~\ref{tab:syst}.
\begin{table*}
\begin{center}
\caption{\label{tab:syst} Summary of the uncertainties in the
determination of \asmz, in units of $10^{-3}$.}
\begin{tabular}{lcccccc}
\toprule
Experimental uncertainty         & \multicolumn{2}{c}{$\pm0.44$}\\
PDF uncertainty                  & \multicolumn{2}{c}{$\pm0.51$}\\
Scale variation uncertainties  & \multicolumn{2}{c}{$\pm0.42$}\\
Matching to fixed order          &        \phantom{+}0\phantom{.00} & $-0.08$\\
Non-perturbative model           & +0.12 & $-0.20$\\
Flavour model                    & +0.40 & $-0.29$\\
QED ISR                          & \multicolumn{2}{c}{$\pm0.14$}\\
N$^4$LL approximation               & \multicolumn{2}{c}{$\pm0.04$}\\
\midrule
Total                            & +0.91 & $-0.88$\\
\bottomrule
\end{tabular}
\end{center}
\end{table*}
 
The goodness of fit is assessed by computing the value of the $\chi^2$
function with the theory predictions evaluated at the measured value
of \asmz{} and with the best-fit values of the non-perturbative parameters
and the QCD scales. In addition to the PDF uncertainties included
in Eq.~(\ref{eq:chi2prof}), all theory uncertainties considered in the
analysis are added as theory nuisance parameters. The computed
$\chi^2$ value is $82$ for $72$ data points, corresponding to a $p$-value of
$0.2$.
 
The upper end of the fit range is varied to test
the stability of the results with respect to missing higher-order
corrections in the matching to fixed order. Lowering the upper
end from $29$~\GeV{} to $22$~\GeV{} shifts the \asmz{} value by $-0.00017$ and
increases the estimated missing higher-order uncertainties from
$0.00042$ to $0.00050$. Raising the upper end to
$40.4$~\GeV{} shifts the \asmz{} value by $+0.00028$ and
increases the estimated missing higher-order uncertainties to
$0.00088$. The shifts in the central value are compatible
with the increase in the missing higher-order uncertainties.
 
The fit range is also varied by excluding the low
transverse-momentum region. The range is narrowed to $5 < \pt <
29$~\GeV, with a spread in the values of $\asmz$ at the level of
$^{+0.00017}_{-0.00010}$, compatible with the increase in the
uncertainty of the fit, from $0.00067$ to $0.00071$.
Since the low transverse-momentum region of $\pt < 5$~\GeV{} is the
most sensitive to the non-perturbative and quark-flavour effects, this
test provides a strong validation of the modelling of these corrections.

\begin{figure}
\centering
\includegraphics[width=\textwidth]{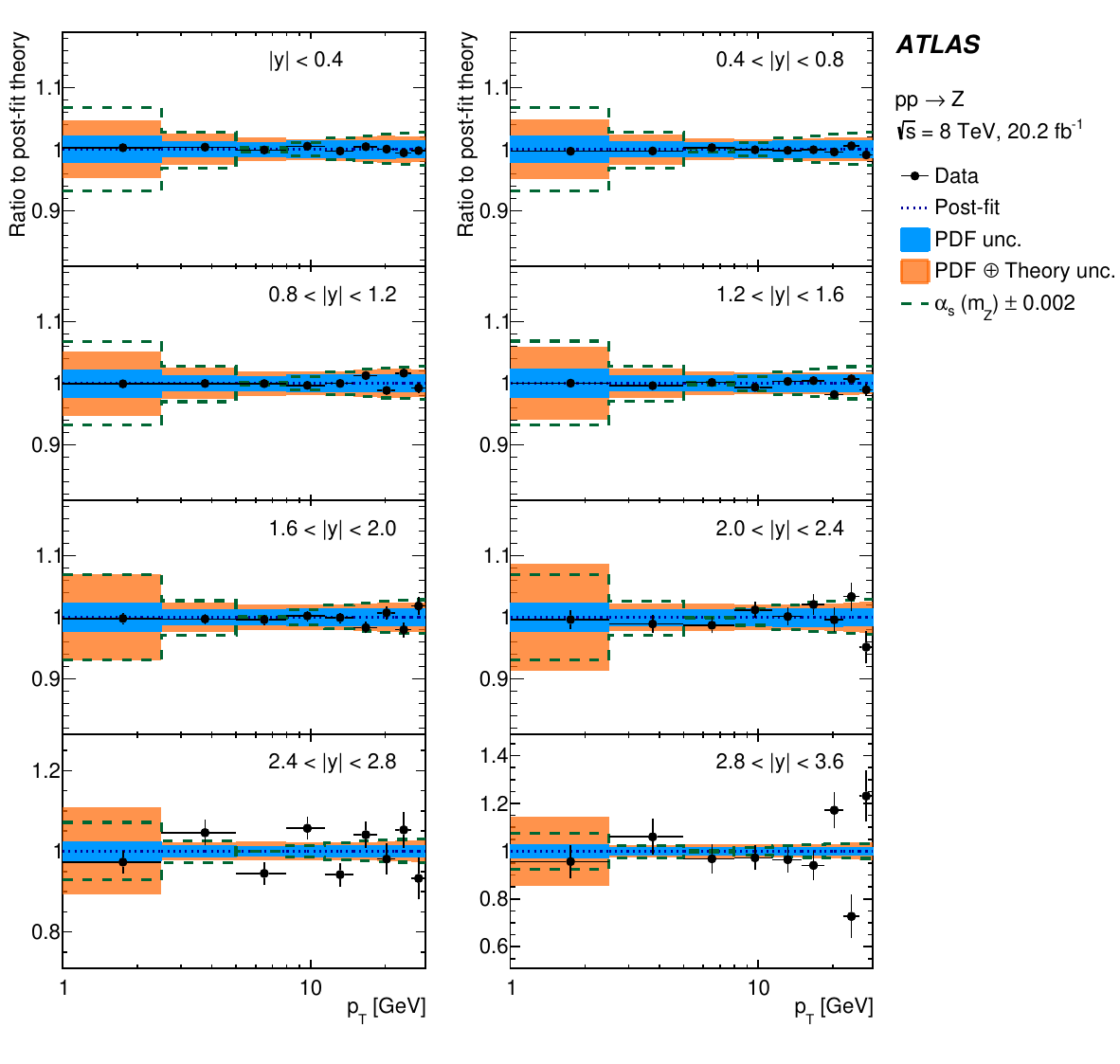}
\caption{Ratios of the measured double-differential cross sections to the post-fit
predictions, both as functions of the transverse momentum and rapidity of the $Z$ boson.
The blue inner band shows the PDF uncertainties of the
predictions pulled and constrained by the fit, and the orange band shows the PDF and
all other unconstrained theoretical uncertainties added in quadrature. The measured cross
sections are corrected by the post-fit pull of the luminosity
uncertainty. The vertical error bars show the experimental
uncertainties of the measurement. The dashed lines show post-fit
predictions in which \asmz{} is varied by $\pm 0.002$ and all other
parameters are kept fixed.
\label{fig:fit}}
\end{figure}

The post-fit predictions are compared with the measured $Z$-boson
transverse-momentum distribution in Figure~\ref{fig:fit}.
The overall change in the normalisation is accounted for by a pull of
the 1.8\% luminosity uncertainty by +1.3~standard deviations.
 
The determination of \asmz{} is repeated at a lower order,
N$^3$LL+N$^3$LO, with the MSHT20, CT18A, NNPDF4.0 and HERAPDF2.0 NNLO
PDF sets. The spread of the fitted values of \asmz{} is $\pm 0.00102$,
driven by the difference between CT18A and NNPDF4.0. While these PDF
sets are not appropriate for the present measurement given their
lower theoretical accuracy, this study provides a conservative
estimate of the residual PDF model dependence of the result,
demonstrating that the achievable accuracy is excellent compared to
that of other methods of extracting \asmz.
At this order, in addition to the Hessian profiling approach, a
simultaneous PDF-fit determination is performed through the numerical
minimisation of the $\chi^2$ in the full multidimensional parameter space
of PDFs, the non-perturbative parameters, and \asmz.
The combined neutral- and charged-current deep inelastic scattering (DIS)
cross-section data from the H1 and ZEUS experiments at the HERA
collider~\cite{H1:2015ubc} are included in the fit, with a minimum
squared four-momentum transfer $Q^2$ of 10~$\GeV^2$, together with the
measured $Z$-boson transverse-momentum distribution.
The value of \asmz{} determined from this fit is $0.11866 \pm 0.00064$,
where the quoted uncertainty is the uncertainty from the fit,
which includes experimental and PDF uncertainties.
The determined value of \asmz{} is in agreement with corresponding
determinations using the Hessian profiling approach at this order, and
the uncertainty is comparable to the uncertainty of the nominal fit.
At N$^3$LL+N$^3$LO, missing higher-order uncertainties estimated with scale variations
amount to $\pm0.00066$. Considering all the other relevant
uncertainties listed in Table~\ref{tab:syst}, the result of this
determination is $\asmz = 0.1187 \pm 0.0010$.

\FloatBarrier
 
\section{Outlook}
 
\begin{figure}
\centering
\includegraphics[width=0.9\textwidth]{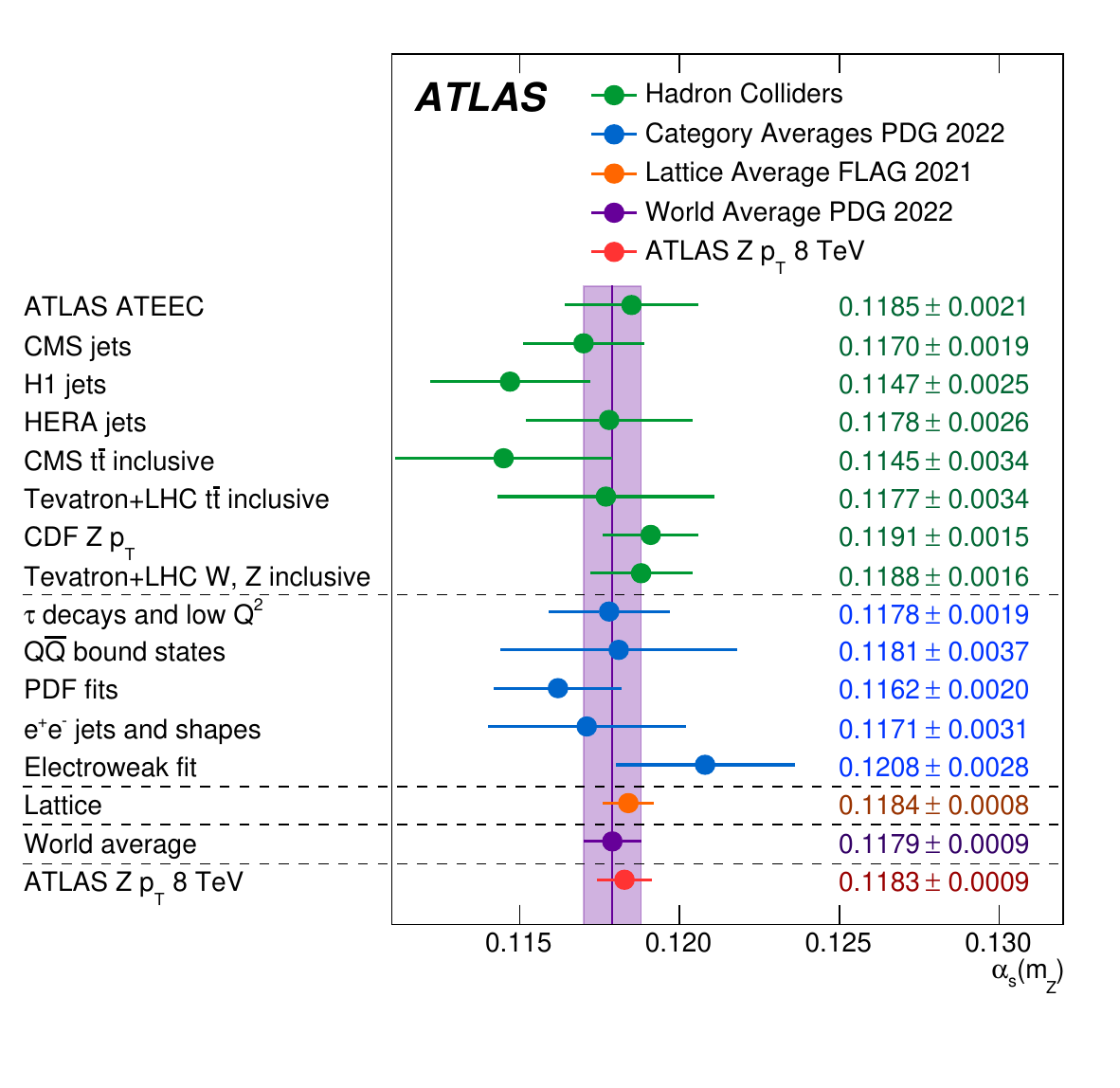}
\caption{Comparison of the determination of \asmz{} from
the \Zboson-boson transverse-momentum distribution (ATLAS $Z$ \pT{} 8~\TeV) with other
determinations at hadron colliders~\cite{STDM-2018-51,CMS-SMP-20-011,H1:2017bml,Britzger:2019kkb,Camarda:2022qdg,dEnterria:2019aat,CMS-TOP-17-001,Klijnsma:2017eqp}, with the PDG category averages~\cite{ParticleDataGroup:2022pth},
with the lattice QCD determination~\cite{Aoki:2021kgd}, and with the PDG world average~\cite{ParticleDataGroup:2022pth}.\label{fig:summary}}
\end{figure}

The coupling constant of the strong force is determined
from a measurement of the transverse-momentum distribution of $Z$
bosons produced at the LHC in 20.2~fb$^{-1}$ of 8~\TeV\ proton--proton collisions and recorded by the ATLAS experiment.
The analysis is based on a semi-inclusive observable at hadron--hadron
colliders, and employs QCD resummed theory predictions.
In contrast to other hadron collider observables, the $Z$-boson
transverse-momentum distribution in the Sudakov region is not included
in PDF fits, thus largely removing the issue of correlation between
this \asmz{} determination and simultaneous determinations of PDFs and the strong-coupling constant.
The measured value of $\asmz = 0.1183 \pm 0.0009$ is compatible with other determinations and with the
world-average value, as illustrated in Figure~\ref{fig:summary}.
 
Among experimental determinations, this is the most precise to
date and the first based on N$^4$LLa+N$^3$LO predictions in perturbative QCD.
This result marks the start of a new era in precision studies of QCD
with the Drell--Yan process. Using this approach, the strong-coupling constant can be
investigated with higher precision and in higher energy regimes with
future larger datasets.

\FloatBarrier
 
\section*{Acknowledgements}


We thank CERN for the very successful operation of the LHC, as well as the
support staff from our institutions without whom ATLAS could not be
operated efficiently.
 
We acknowledge the support of
ANPCyT, Argentina;
YerPhI, Armenia;
ARC, Australia;
BMWFW and FWF, Austria;
ANAS, Azerbaijan;
CNPq and FAPESP, Brazil;
NSERC, NRC and CFI, Canada;
CERN;
ANID, Chile;
CAS, MOST and NSFC, China;
Minciencias, Colombia;
MEYS CR, Czech Republic;
DNRF and DNSRC, Denmark;
IN2P3-CNRS and CEA-DRF/IRFU, France;
SRNSFG, Georgia;
BMBF, HGF and MPG, Germany;
GSRI, Greece;
RGC and Hong Kong SAR, China;
ISF and Benoziyo Center, Israel;
INFN, Italy;
MEXT and JSPS, Japan;
CNRST, Morocco;
NWO, Netherlands;
RCN, Norway;
MEiN, Poland;
FCT, Portugal;
MNE/IFA, Romania;
MESTD, Serbia;
MSSR, Slovakia;
ARRS and MIZ\v{S}, Slovenia;
DSI/NRF, South Africa;
MICINN, Spain;
SRC and Wallenberg Foundation, Sweden;
SERI, SNSF and Cantons of Bern and Geneva, Switzerland;
MOST, Taiwan;
TENMAK, T\"urkiye;
STFC, United Kingdom;
DOE and NSF, United States of America.
In addition, individual groups and members have received support from
BCKDF, CANARIE, Compute Canada and CRC, Canada;
PRIMUS 21/SCI/017 and UNCE SCI/013, Czech Republic;
COST, ERC, ERDF, Horizon 2020 and Marie Sk{\l}odowska-Curie Actions, European Union;
Investissements d'Avenir Labex, Investissements d'Avenir Idex and ANR, France;
DFG and AvH Foundation, Germany;
Herakleitos, Thales and Aristeia programmes co-financed by EU-ESF and the Greek NSRF, Greece;
BSF-NSF and MINERVA, Israel;
Norwegian Financial Mechanism 2014-2021, Norway;
NCN and NAWA, Poland;
La Caixa Banking Foundation, CERCA Programme Generalitat de Catalunya and PROMETEO and GenT Programmes Generalitat Valenciana, Spain;
G\"{o}ran Gustafssons Stiftelse, Sweden;
The Royal Society and Leverhulme Trust, United Kingdom.
 
The crucial computing support from all WLCG partners is acknowledged gratefully, in particular from CERN, the ATLAS Tier-1 facilities at TRIUMF (Canada), NDGF (Denmark, Norway, Sweden), CC-IN2P3 (France), KIT/GridKA (Germany), INFN-CNAF (Italy), NL-T1 (Netherlands), PIC (Spain), ASGC (Taiwan), RAL (UK) and BNL (USA), the Tier-2 facilities worldwide and large non-WLCG resource providers. Major contributors of computing resources are listed in Ref.~\cite{ATL-SOFT-PUB-2023-001}.


\clearpage
\section{Methods}
\label{sec:methods}
 
\textbf{Modelling of non-perturbative effects.}
Determinations of \asmz{} are affected by non-perturbative power
corrections of the type $\Lambda_{\textrm{QCD}}^p/Q^p$, where
$\Lambda_{\textrm{QCD}}$ is the non-perturbative scale of QCD and $Q$
is the order of magnitude of the momentum transfer in the process.
Their impact strongly depends on the value of the power $p$ for the
given process used to determine \asmz.
Non-perturbative QCD effects~\cite{Collins:1984kg,Davies:1984sp,Ladinsky:1993zn,Ellis:1997sc,Landry:1999an,Qiu:2000hf,Kulesza:2002rh,Konychev:2005iy,Guzzi:2013aja,Collins:2014jpa,Wei:2020glg},
are expected to be quadratically suppressed for the Drell--Yan \pt{}
distribution at large \pt, or, equivalently,
in the limit of small $b$~\cite{Tafat:2001in}, thanks to the azimuthal symmetry of
the intrinsic transverse-momentum smearing of partons~\cite{FerrarioRavasio:2020guj,Caola:2021kzt}.
In the small \pt{} region the non-perturbative corrections are
expected to become linear below a given
scale~\cite{Vladimirov:2020umg,Collins:2014jpa}, which is estimated of
the order of 1~\GeV in Ref.~\cite{Schweitzer:2012hh}.
 
In the case of the \Zboson-boson \qt-resummed predictions used in this
analysis, the Sudakov form factor is singular in the region of transverse momenta
of the order of the QCD scale
$\Lambda_{\textrm{QCD}}$.
The singular behaviour of the perturbative form factor is removed
by using the so-called $b_*$~\cite{Collins:1981va,Collins:1984kg}
regularisation procedure, in which the dependence of the Sudakov form
factor on the impact parameter $b$ is frozen before reaching the
singular point by performing the replacement $b^2 \to b_*^2 = b^2 b_{\textrm{lim}}^2/(
b^2 + b_{\textrm{lim}}^2)$. A default value of
$b_{\textrm{lim}} = 2$~$\GeV^{-1}$ is used in the calculation.
Non-perturbative QCD effects are included in this analysis with a non-perturbative form
factor~\cite{Collins:1984kg,Collins:2014jpa}:
\begin{eqnarray}
\label{eq:nonpqcd_1}
S_{\textrm{NP}}(b) = \exp \left[ -g_j(b) -g_K(b) \log{\frac{m_{\ell\ell}^2}{Q_0^2}} \right] \,
\end{eqnarray}
with
\begin{eqnarray}
g_{j}(b) = \frac{g\,b^2}{\sqrt{1+\lambda\,b^2}}\ + \textrm{sign}(q)   \left( 1
-\exp \left[-|q|\,b^{4} \right] \right)\\
\label{eq:nonpqcd_3}
g_{K}(b) = g_0 \left(1 - \exp \left[ - \frac{C_F
\alphas(b_0/b_*) b^2}{\pi g_0\, b_{\textrm{lim}}^2} \right] \right) \,,
\end{eqnarray}
 
where $b_0 = 2\textrm{e}^{-\gamma_E}$, and $\gamma_E$ is the Euler number.
The $g$ and $q$ parameters represent the leading quadratic and
quartic terms which are dominant in the region of moderate \pt{} of
$4$--$10$~\GeV, where the sensitivity to \asmz{} is maximal, and they
are left free in the fit. The parameter $\lambda$ controls the scale
of transition from quadratic (Gaussian) to linear (exponential)
behaviour of the non-perturbative primordial $k_\mathrm{T}$. It is set to $1$~$\GeV^2$ and varied when
assessing uncertainties of the non-perturbative model.
The parameter $g_0$ controls the asymptotic
behaviour of the non-perturbative form factor at very small \pt, in a region where
the measured cross section and the determined value of \asmz{} have
very little sensitivity. It is set to
$0.3$~\cite{Collins:2014jpa} and varied when
assessing uncertainties of the non-perturbative model.
The parameters $b_{\textrm{lim}}$ and $Q_0$ represent respectively the
scale at which the running of \as{} is frozen, and the starting scale
at which the non-perturbative form factor is parameterised by the function $g_{j}(b)$.
Changes in these parameters should be completely reabsorbed by
changes in the functions $g_{K}(b)$ and $g_{j}(b)$, provided they are
flexible enough. Variations of $b_{\textrm{lim}}$ and of $Q_0$ are
performed to assess the uncertainty related to the choice of
parameterisation in Eqs.~(\ref{eq:nonpqcd_1})--(\ref{eq:nonpqcd_3}).
 
The value of $g$ determined in the nominal fit
is $g = 0.54 \pm 0.04$~$\GeV^2$ and its correlation with \asmz{} is
$-0.6$. The value of $q$ determined in the fit
is $q = -0.06 \pm 0.04$~$\GeV^4$ and its correlation with \asmz{} is
$+0.4$. The correlation between $g$ and $q$ is $-0.7$.
Uncertainties in the modelling of the non-perturbative form factor
are estimated with variations of the parameters $b_{\textrm{lim}}$,
$Q_0$, $g_0$, and $\lambda$.
Variations of $b_{\textrm{lim}}$ in the range $1.5$--$2.5$~$\GeV^{-1}$
yield \asmz{} variations of $^{+0.00012}_{-0.00020}$.
Variations of $Q_0$ in the range $0.5$--$2$~\GeV{} yield \asmz{} variations of $^{+0.00006}_{-0.00002}$.
Variations of $g_0$ in the range $0.1$--$0.5$ yield \asmz{} variations
which are at the level of $\pm0.00002$.
Variations of $\lambda$ in the range $0.5$--$2$~$\GeV^2$ yield \asmz{} variations of $^{+0.00011}_{-0.00019}$.
The envelope of these variations is $^{+0.00012}_{-0.00020}$, which is
used as an estimate of the uncertainty in the non-perturbative
model.
 
\begin{figure}
\centering
\subfigure[]{\includegraphics[width=0.45\textwidth]{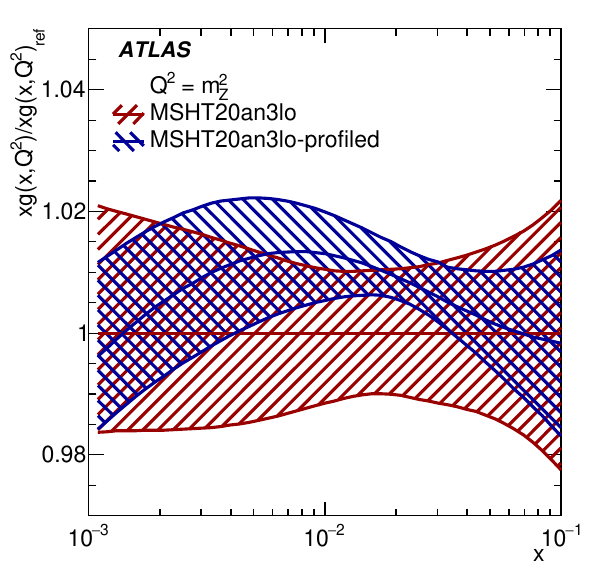}}
\subfigure[]{\includegraphics[width=0.45\textwidth]{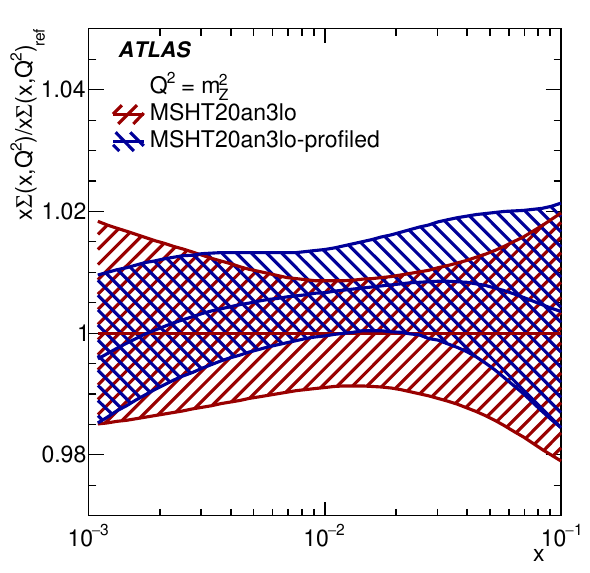}}
\caption{Ratios of the profiled (a) gluon and (b) sea-quark PDFs to
their initial values, at the scale $q^2 = m_Z^2$. The error bands
represent the 68\% confidence level.\label{fig:pdfs}}
\end{figure}

\textbf{PDF profiling.}
Pulls and constraints on the nuisance parameters associated with the PDF
uncertainties in Eq.~(\ref{eq:chi2prof}) can be reinterpreted in the
PDF space through a Hessian profiling
procedure~\cite{HERAFitterdevelopersTeam:2015cre}. Such a
reinterpretation provides valuable information about the sensitivity of the
measured cross sections to the PDFs. The largest observed effects are
on the gluon and sea-quark PDFs, which are shown in Figure~\ref{fig:pdfs}.
 
\textbf{Fits with NNLO PDFs.}
At order N$^4$LLa+N$^3$LO, only one N$^3$LO PDF set is currently available,
namely the MSHT20aN$^3$LO~\cite{McGowan:2022nag} PDF set. In order to
study the dependence of the results on the choice of PDF set, fits
are performed at a lower order, N$^3$LL+N$^3$LO, using NNLO PDF sets.
Table~\ref{tab:nnlo} shows results of fits with various PDF
sets. At this order, the spread observed in the values of \asmz{}
extracted with different PDF sets is $\pm 0.00102$, which is driven by
the difference between the NNPDF4.0 and CT18A PDF sets.
\begin{table*}
\begin{center}
\caption{\label{tab:nnlo} Summary of N$^3$LL fits with NNLO PDFs.}
\begin{tabular}{lcccc}
\toprule
PDF set & \asmz & PDF uncertainty & $g$ [$\GeV^2$] & $q$ [$\GeV^4$]\\
\midrule
MSHT20~\cite{Bailey:2020ooq} & 0.11839 & 0.00040 & 0.44 & $-0.07$ \\
NNPDF4.0~\cite{NNPDF:2021njg} & 0.11779 & 0.00024 & 0.50 & $-0.08$ \\
CT18A~\cite{Hou:2019efy}     & 0.11982 & 0.00050 & 0.36 & $-0.03$ \\
HERAPDF2.0~\cite{H1:2015ubc}  & 0.11890 & 0.00027 & 0.40 & $-0.04$ \\
\bottomrule
\end{tabular}
\end{center}
\end{table*}
 
The determination of \asmz{} from the transverse-momentum distribution
of $Z$ bosons is particularly sensitive to the gluon PDF. The PDF
determinations at NNLO are affected by significant tension between
the low-$x$ and high-$x$ gluon PDFs, which is ascribed to tensions between
observables sensitive to the gluon PDF, such as those for inclusive deep inelastic
scattering (DIS) at the HERA collider, hadron-collider jet measurements,
top-quark pair production, and $Z$-boson \pt{} measurements in the high
transverse-momentum region.
 
In order to investigate the effect of these tensions on the
determination of \asmz{} at N$^3$LL, fits are performed which also include
the combined neutral- and charged-current DIS cross-section data from the H1 and ZEUS
experiments at the HERA collider~\cite{H1:2015ubc}, with
a minimum $Q^2$ value of 10~$\GeV^2$, together with the measured
$Z$-boson transverse-momentum cross sections.
The HERA data are already included in all PDF fits, they are included
again here for the purpose of lessening the impact of other
datasets on the gluon PDF. After the inclusion of HERA data in the fit,
all PDF sets yield values of \asmz{} consistent with the initial
results, except for CT18A, which is shifted by $-0.00166$
compared to a PDF uncertainty of $\pm 0.00050$, and the half-envelope
of $\asmz$ values for the PDF sets considered is reduced to $\pm
0.00016$.
 
The approximate N$^3$LO PDF fit of MSHT20, which is used for the nominal
result, largely removes the tension in the gluon PDF, as indicated by
the significant improvement in the $\chi^2$ associated with the
$Z$-boson \pt{} measurement in the high transverse-momentum region and
with inclusive DIS at the HERA collider, compared to the NNLO fit~\cite{McGowan:2022nag}.
These observations suggest that the spread in \asmz{} when using different
PDF sets at NNLO is not representative of the true PDF uncertainty at
N$^3$LO. However, further studies to verify the robustness of the
estimate of the PDF uncertainties at N$^3$LO in the MSHT20 analysis
will be possible when other PDF determinations at this order become
available.
 
\textbf{Combined fits of \asmz{} and PDFs.}
Determinations of $\asmz$ at hadron colliders are exposed to possible
biases unless the PDFs are determined simultaneously along with
$\asmz$~\cite{Forte:2020pyp}.
The Hessian profiling employed in this analysis provides an
approximation to a PDF determination which relies on the accuracy of
the quadratic approximation around the
minimum~\cite{Paukkunen:2014zia}. In the nominal fit of \asmz{} at
N$^4$LLa+N$^3$LO, pulls and constraints on the nuisance parameters
associated with the PDF uncertainties are below one standard deviation and $30\%$,
respectively, indicating that the new minimum of the profiled PDFs is
close to the original minimum, which gives confidence in the validity
of the quadratic approximation.
 
A simultaneous determination of \asmz, the PDFs, and the non-perturbative
parameters through the numerical minimisation of the $\chi^2$ in the
full multidimensional parameters space~\cite{Agarwal_Ceres_Solver_2022} is
performed at N$^3$LL+N$^3$LO, with PDFs evolved at NNLO.
The combined neutral- and charged-current DIS cross-section data from the H1 and ZEUS
experiments at the HERA collider~\cite{H1:2015ubc} are included, with
a minimum squared four-momentum transfer $Q^2$ of 10~$\GeV^2$, corresponding
to 1016 data points, together with the measured $Z$-boson
transverse-momentum cross sections.
 
The light-quark coefficient functions of the DIS cross sections are
calculated in the $\overline{MS}$ scheme~\cite{Weinberg:1973xwm}, and
with the renormalisation and factorisation scales set to the squared four-momentum transfers $Q^2$.
The heavy quarks $c$ and $b$ are generated dynamically, and the
corresponding coefficient functions for the neutral-current processes
with $\gamma^*$ exchange are calculated in the general-mass
variable-flavour-number
scheme~\cite{Thorne:1997ga,Thorne:2006qt,Thorne:2012az}, with up to
five active quark flavours.
The charm-quark mass is set to $m_c = 1.43$~\GeV, and the bottom-quark mass to $m_b
= 4.50$~\GeV~\cite{H1:2015ubc}. For the charged-current processes, the
heavy quarks are treated as massless.
 
The PDFs for the gluon, valence $u$- and $d$-quark, and $\bar{u}$- and $\bar{d}$-quark densities
are parameterised at the input scale
$Q^2_0=1.9$~GeV$^2$ with the parameterisation in Ref.~\cite{Bonvini:2019wxf}.
The contribution of the $s$-quark density is taken to be proportional to the
$\bar{d}$-quark density by setting $x\bar{s}(x) = r_s x\bar{d}(x)$,
with $r_s=0.67$.
 
\begin{figure}
\centering
\includegraphics[width=0.7\textwidth]{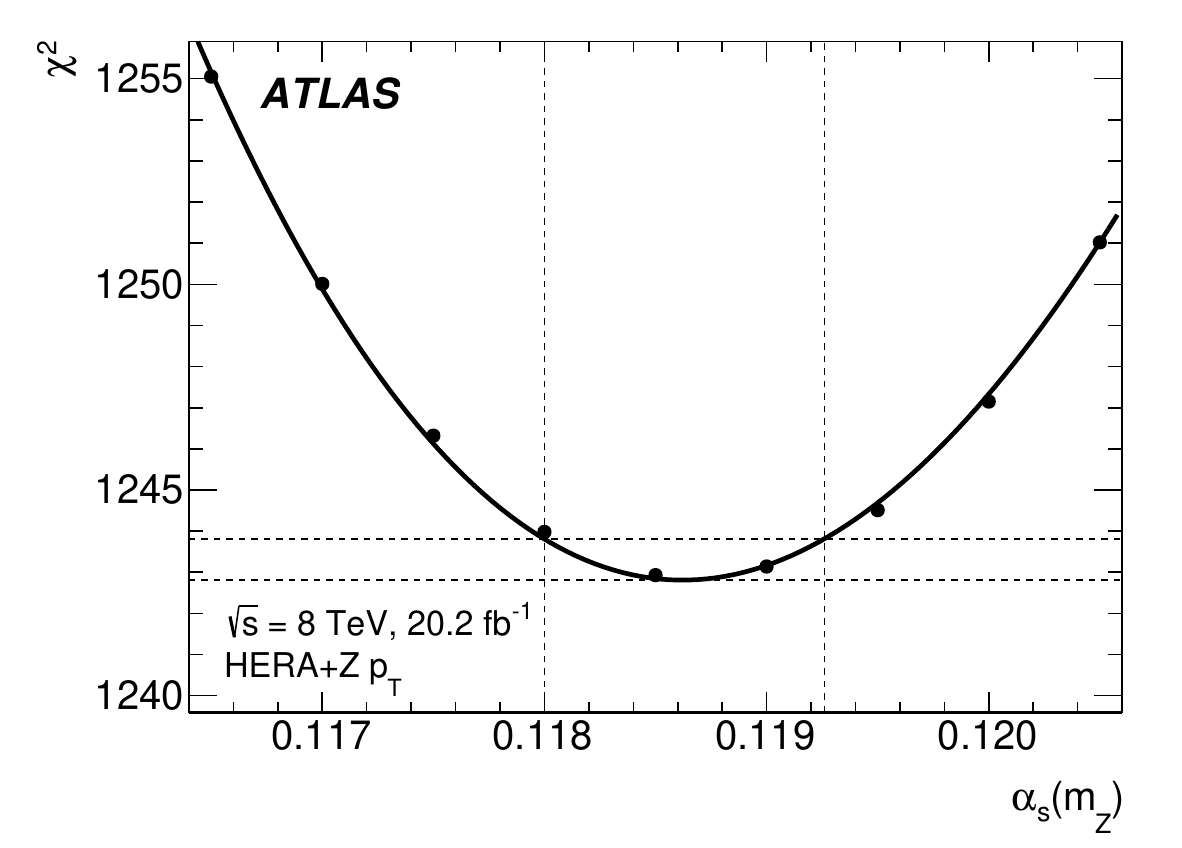}
\caption{Values of the $\chi^2$ function for the determination
of \asmz{} from a simultaneous fit of PDFs and non-perturbative parameters at N$^3$LL+N$^3$LO, with PDFs evolved at NNLO.
The horizontal dashed lines indicate the minimum value of
the $\chi^2$ and the minimum $\chi^2$ plus one. The vertical dashed
lines show the intersections of the line at the minimum $\chi^2$ plus
one with the $\chi^2$ function, which correspond to the \asmz{}
uncertainties from the fit.
\label{fig:herascan}}
\end{figure}

\begin{figure}
\centering
\includegraphics[width=0.7\textwidth]{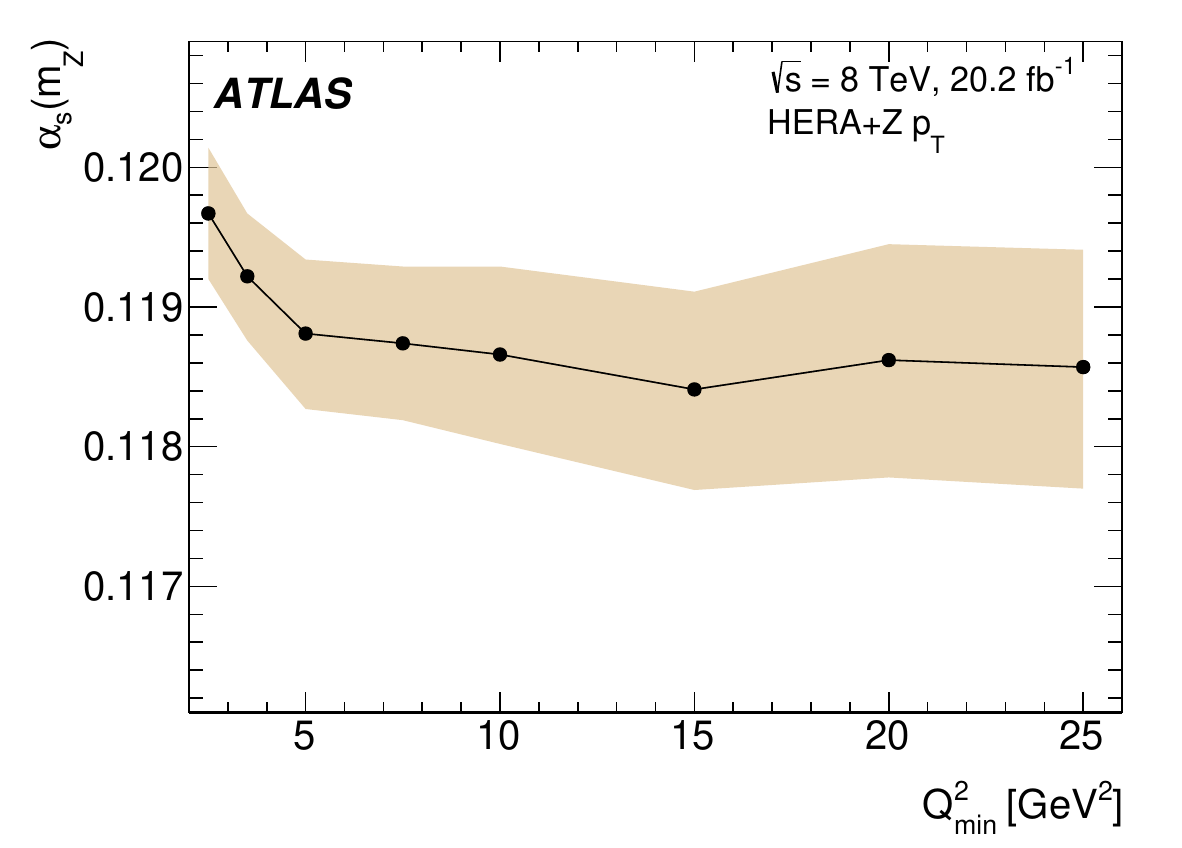}
\caption{Dependence of \asmz{} on the minimum squared four-momentum transfer
$Q^2$ of the HERA data from a simultaneous fit of PDFs and non-perturbative
parameters at N$^3$LL+N$^3$LO, with PDFs evolved at NNLO. The shaded area shows the uncertainty from
the fit\label{fig:qminscan}}
\end{figure}

Fits are performed at fixed values of \asmz, and the fitted value
of \asmz{} is determined from a quadratic interpolation of the $\chi^2$
as a function of \asmz, as shown in Figure~\ref{fig:herascan}.
The determined value of \asmz{} is $0.11866 \pm 0.00064$, where the quoted
uncertainty is the uncertainty from the fit, which includes
experimental and PDF uncertainties. The value of \asmz{} is in
agreement with corresponding determinations using the Hessian profiling
approach at this order, as shown in Table~\ref{tab:nnlo}, and the
uncertainty is comparable to the uncertainty of $\pm0.00067$ in the nominal fit.
The dependence of \asmz on the minimum squared four-momentum transfer $Q^2$
of the HERA data is studied in the range from $2.5$~$\GeV^2$ to
$25$~$\GeV^2$, as shown in Figure~\ref{fig:qminscan}. No significant
dependence is observed above $5$~$\GeV^2$.
 
\FloatBarrier
 
\textbf{Sudakov subleading higher-order corrections.}
Additional fits are performed which differ from the
nominal fit for subleading higher-order corrections in the Sudakov
form factor.
In Ref.~\cite{Billis:2019evv}, different procedures for the computation
of the Sudakov form factor are discussed, including analytic and
numerical solutions. Subleading corrections in the definition of the
Sudakov form factor and in the running of \as{} are tested with fits
where the Sudakov form factor is evaluated with a numerical
integration, and $\as(Q)$ in the Sudakov form factor is obtained from the exact numerical
renormalisation-group-equation invariant solution for the running
of \as{}. The effects of scale variations are estimated with the
methodology proposed in Ref.~\cite{Bertone:2022sso}.
For these fits, the midpoint and half envelope of the \asmz{} values from the scale variations
yield $\asmz = 0.11832 \pm 0.00029$.
Fits where the hard-collinear coefficients are evolved according to the CSS
scheme~\cite{Collins:1984kg} yield a value of $\asmz = 0.11872$ for the
central value of the scales. In all the cases considered, the
inclusion of subleading higher-order corrections is covered by the
estimate of missing higher-order corrections based on scale
variations, so no additional uncertainty is considered.

\printbibliography
 
\clearpage
\input{atlas_authlist}

\end{document}

%% file: atlas_authlist.tex
 
\begin{flushleft}
\hypersetup{urlcolor=black}
{\Large The ATLAS Collaboration}

\bigskip

\AtlasOrcid[0000-0002-6665-4934]{G.~Aad}$^\textrm{\scriptsize 102}$,
\AtlasOrcid[0000-0002-5888-2734]{B.~Abbott}$^\textrm{\scriptsize 120}$,
\AtlasOrcid[0000-0002-1002-1652]{K.~Abeling}$^\textrm{\scriptsize 55}$,
\AtlasOrcid[0000-0001-5763-2760]{N.J.~Abicht}$^\textrm{\scriptsize 49}$,
\AtlasOrcid[0000-0002-8496-9294]{S.H.~Abidi}$^\textrm{\scriptsize 29}$,
\AtlasOrcid[0000-0002-9987-2292]{A.~Aboulhorma}$^\textrm{\scriptsize 35e}$,
\AtlasOrcid[0000-0001-5329-6640]{H.~Abramowicz}$^\textrm{\scriptsize 151}$,
\AtlasOrcid[0000-0002-1599-2896]{H.~Abreu}$^\textrm{\scriptsize 150}$,
\AtlasOrcid[0000-0003-0403-3697]{Y.~Abulaiti}$^\textrm{\scriptsize 117}$,
\AtlasOrcid[0000-0003-0762-7204]{A.C.~Abusleme~Hoffman}$^\textrm{\scriptsize 137a}$
% ...
\bigskip
\\

% ...
$^{29}$Physics Department, Brookhaven National Laboratory, Upton NY; United States of America.\\
% ...
$^{35}$$^{(a)}$Facult\'e des Sciences Ain Chock, R\'eseau Universitaire de Physique des Hautes Energies - Universit\'e Hassan II, Casablanca;$^{(b)}$Facult\'{e} des Sciences, Universit\'{e} Ibn-Tofail, K\'{e}nitra;$^{(c)}$Facult\'e des Sciences Semlalia, Universit\'e Cadi Ayyad, LPHEA-Marrakech;$^{(d)}$LPMR, Facult\'e des Sciences, Universit\'e Mohamed Premier, Oujda;$^{(e)}$Facult\'e des sciences, Universit\'e Mohammed V, Rabat;$^{(f)}$Institute of Applied Physics, Mohammed VI Polytechnic University, Ben Guerir; Morocco.\\
% ...
$^{49}$Fakult\"{a}t Physik , Technische Universit{\"a}t Dortmund, Dortmund; Germany.\\
% ...
$^{55}$II. Physikalisches Institut, Georg-August-Universit\"{a}t G\"ottingen, G\"ottingen; Germany.\\
% ...
$^{102}$CPPM, Aix-Marseille Universit\'e, CNRS/IN2P3, Marseille; France.\\
% ...
$^{117}$Department of Physics, New York University, New York NY; United States of America.\\
% ...
$^{120}$Homer L. Dodge Department of Physics and Astronomy, University of Oklahoma, Norman OK; United States of America.\\
% ...
$^{137}$$^{(a)}$Departamento de F\'isica, Pontificia Universidad Cat\'olica de Chile, Santiago;$^{(b)}$Millennium Institute for Subatomic physics at high energy frontier (SAPHIR), Santiago;$^{(c)}$Instituto de Investigaci\'on Multidisciplinario en Ciencia y Tecnolog\'ia, y Departamento de F\'isica, Universidad de La Serena;$^{(d)}$Universidad Andres Bello, Department of Physics, Santiago;$^{(e)}$Instituto de Alta Investigaci\'on, Universidad de Tarapac\'a, Arica;$^{(f)}$Departamento de F\'isica, Universidad T\'ecnica Federico Santa Mar\'ia, Valpara\'iso; Chile.\\
% ...
$^{150}$Department of Physics, Technion, Israel Institute of Technology, Haifa; Israel.\\
$^{151}$Raymond and Beverly Sackler School of Physics and Astronomy, Tel Aviv University, Tel Aviv; Israel.\\
% ...
$^{a}$ Also Affiliated with an institute covered by a cooperation agreement with CERN.\\
$^{b}$ Also at An-Najah National University, Nablus; Palestine.\\
$^{c}$ Also at APC, Universit\'e Paris Cit\'e, CNRS/IN2P3, Paris; France.\\
$^{d}$ Also at Borough of Manhattan Community College, City University of New York, New York NY; United States of America.\\
$^{e}$ Also at Center for High Energy Physics, Peking University; China.\\
$^{f}$ Also at Center for Interdisciplinary Research and Innovation (CIRI-AUTH), Thessaloniki; Greece.\\
% ...

\end{flushleft}


%% file: ANA-STDM-2023-01-PAPER.bib
@Article{STDM-2018-05,
    author         = "{ATLAS Collaboration}",
    title = "{A precise measurement of the Z-boson double-differential transverse momentum and rapidity distributions in the full phase space of the decay leptons with the ATLAS experiment at $\sqrt s$ = 8 TeV}",
    year           = "2023",
    reportNumber   = "CERN-EP-2023-171",
    eprint = "2309.09318",
    archivePrefix  = "arXiv",
    primaryClass   = "hep-ex",
}

@article{ParticleDataGroup:2020ssz,
    author = "Zyla, P. A. and others",
    collaboration = "Particle Data Group",
    title = "{Review of Particle Physics}",
    doi = "10.1093/ptep/ptaa104",
    journal = "PTEP",
    volume = "2020",
    number = "8",
    pages = "083C01",
    year = "2020"
}

@inbook{doi:10.1142/9789813238053_0007,
  author = { Gavin   Salam },
  editor = "Levy, Aharon and Forte, Stefano and Ridolfi, Giovanni",
  title = {The strong coupling: a theoretical perspective},
  booktitle = "{From My Vast Repertoire ...}: {Guido Altarelli's Legacy}",
  chapter = {7},
  pages = "101--121",
  year = "2019",
  doi = {10.1142/9789813238053_0007},
  URL = {https://www.worldscientific.com/doi/abs/10.1142/9789813238053_0007},
  eprint = "1712.05165",
  archivePrefix = "arXiv",
  primaryClass = "hep-ph",
  reportNumber = "CERN-TH-2017-268"
}

@article{Aoki:2021kgd,
    author = "Aoki, Y. and others",
    title = "{FLAG Review 2021}",
    eprint = "2111.09849",
    archivePrefix = "arXiv",
    primaryClass = "hep-lat",
    reportNumber = "CERN-TH-2021-191, JLAB-THY-21-3528, FERMILAB-PUB-21-620-SCD-T",
    doi = "10.1140/epjc/s10052-022-10536-1",
    journal = "Eur. Phys. J. C",
    volume = "82",
    number = "10",
    pages = "869",
    year = "2022"
}

@article{Haller:2018nnx,
    collaboration = "GFitter Group",
    author = "{GFitter Group, J. Haller et al.}",
    title = "{Update of the global electroweak fit and constraints on two-Higgs-doublet models}",
    eprint = "1803.01853",
    archivePrefix = "arXiv",
    primaryClass = "hep-ph",
    doi = "10.1140/epjc/s10052-018-6131-3",
    journal = "Eur. Phys. J. C",
    volume = "78",
    number = "8",
    pages = "675",
    year = "2018"
}

@article{Baikov:2008jh,
    author = "Baikov, P. A. and Chetyrkin, K. G. and Kuhn, Johann H.",
    title = {Order ${\ensuremath{\alpha}}_{s}^{4}$ QCD Corrections to $Z$ and $\ensuremath{\tau}$ Decays},
    eprint = "0801.1821",
    archivePrefix = "arXiv",
    primaryClass = "hep-ph",
    reportNumber = "SFB-CPP-08-04, TTP08-01",
    doi = "10.1103/PhysRevLett.101.012002",
    journal = "Phys. Rev. Lett.",
    volume = "101",
    pages = "012002",
    year = "2008"
}

@article{Klijnsma:2017eqp,
    author = {Klijnsma, Thomas and Bethke, Siegfried and Dissertori, G\"unther and Salam, Gavin P.},
    title = "{Determination of the strong coupling constant $\alpha_s(m_Z)$ from measurements of the total cross section for top-antitop-quark production}",
    eprint = "1708.07495",
    archivePrefix = "arXiv",
    primaryClass = "hep-ph",
    reportNumber = "CERN-TH-2017-183",
    doi = "10.1140/epjc/s10052-017-5340-5",
    journal = "Eur. Phys. J. C",
    volume = "77",
    number = "11",
    pages = "778",
    year = "2017"
}

@article{dEnterria:2019aat,
    author = "d'Enterria, David and Poldaru, Andres",
    title = "{Extraction of the strong coupling $\alpha_{s}$(m$_{Z}$) from a combined NNLO analysis of inclusive electroweak boson cross sections at hadron colliders}",
    eprint = "1912.11733",
    archivePrefix = "arXiv",
    primaryClass = "hep-ph",
    doi = "10.1007/JHEP06(2020)016",
    journal = "JHEP",
    volume = "06",
    pages = "016",
    year = "2020"
}

@article{Boughezal:2017nla,
    author = "Boughezal, Radja and Guffanti, Alberto and Petriello, Frank and Ubiali, Maria",
    title = "{The impact of the LHC Z-boson transverse momentum data on PDF determinations}",
    eprint = "1705.00343",
    archivePrefix = "arXiv",
    primaryClass = "hep-ph",
    reportNumber = "CAVENDISH-HEP-17-05",
    doi = "10.1007/JHEP07(2017)130",
    journal = "JHEP",
    volume = "07",
    pages = "130",
    year = "2017"
}

@article{Ball:2018iqk,
    author = "Ball, Richard D. and Carrazza, Stefano and Del Debbio, Luigi and Forte, Stefano and Kassabov, Zahari and Rojo, Juan and Slade, Emma and Ubiali, Maria",
    collaboration = "NNPDF",
    title = "{Precision determination of the strong coupling constant within a global PDF analysis}",
    eprint = "1802.03398",
    archivePrefix = "arXiv",
    primaryClass = "hep-ph",
    reportNumber = "CAVENDISH-HEP-18-02, CERN-TH-2018-024, Edinburgh 2017/19, Nikhef/2017-046, OUTP-17-14P, DAMTP-2018-3, TIF-UNIMI-2018-1, EDINBURGH-2017-19, NIKHEF-2017-046",
    doi = "10.1140/epjc/s10052-018-5897-7",
    journal = "Eur. Phys. J. C",
    volume = "78",
    number = "5",
    pages = "408",
    year = "2018"
}

@article{Sudakov:1954sw,
    author = "Sudakov, V. V.",
    title = "{Vertex parts at very high-energies in quantum electrodynamics}",
    journal = "Sov. Phys. JETP",
    volume = "3",
    pages = "65--71",
    year = "1956"
}

@article{Drell:1970wh,
    author = "Drell, S. D. and Yan, Tung-Mow",
    title = {Massive Lepton-Pair Production in Hadron-Hadron Collisions at High Energies},
    reportNumber = "SLAC-PUB-0755",
    doi = "10.1103/PhysRevLett.25.316",
    journal = "Phys. Rev. Lett.",
    volume = "25",
    pages = "316--320",
    year = "1970",
    related        = "Drell:1970wh-err",
    relatedstring  = "Erratum:",
}

@article{Camarda:2021ict,
    author = "Camarda, Stefano and Cieri, Leandro and Ferrera, Giancarlo",
    title = {Drell--Yan lepton-pair production: ${q}_{T}$ resummation at ${\mathrm{N}}^{3}\mathrm{LL}$ accuracy and fiducial cross sections at ${\mathrm{N}}^{3}\mathrm{LO}$},
    eprint = "2103.04974",
    archivePrefix = "arXiv",
    primaryClass = "hep-ph",
    doi = "10.1103/PhysRevD.104.L111503",
    journal = "Phys. Rev. D",
    volume = "104",
    number = "11",
    pages = "L111503",
    year = "2021"
}

@article{Chen:2022cgv,
    author = "Chen, Xuan and Gehrmann, Thomas and Glover, E. W. N. and Huss, Alexander and Monni, Pier Francesco and Re, Emanuele and Rottoli, Luca and Torrielli, Paolo",
    title = {Third-Order Fiducial Predictions for Drell--Yan Production at the LHC},
    eprint = "2203.01565",
    archivePrefix = "arXiv",
    primaryClass = "hep-ph",
    reportNumber = "CERN-TH-2022-023, IPPP/22/08, LAPTH-009/22, KA-TP-03-2022,
  P3H-22-022, ZU-TH 07/22",
    doi = "10.1103/PhysRevLett.128.252001",
    journal = "Phys. Rev. Lett.",
    volume = "128",
    number = "25",
    pages = "252001",
    year = "2022"
}

@article{Neumann:2022lft,
    author = "Neumann, Tobias and Campbell, John",
    title = {Fiducial Drell--Yan production at the LHC improved by transverse-momentum resummation at ${\mathrm{N}}^{4}\mathrm{L}{\mathrm{L}}_{\mathrm{p}}+{\mathrm{N}}^{3}\mathrm{LO}$},
    eprint = "2207.07056",
    archivePrefix = "arXiv",
    primaryClass = "hep-ph",
    reportNumber = "FERMILAB-PUB-22-528-T",
    doi = "10.1103/PhysRevD.107.L011506",
    journal = "Phys. Rev. D",
    volume = "107",
    number = "1",
    pages = "L011506",
    year = "2023"
}

@article{Collins:1984kg,
    author = "Collins, John C. and Soper, Davison E. and Sterman, George F.",
    title = "{Transverse Momentum Distribution in Drell--Yan Pair and W and Z Boson Production}",
    reportNumber = "CERN-TH-3923",
    doi = "10.1016/0550-3213(85)90479-1",
    journal = "Nucl. Phys. B",
    volume = "250",
    pages = "199--224",
    year = "1985"
}

@article{Davies:1984sp,
title = {Drell--Yan cross sections at small transverse momentum},
author = "Davies, C. T. H. and Webber, B. R. and Stirling, W. James",
journal = "Nucl. Phys. B",
volume = {256},
pages = {413-433},
year = {1985},
doi = {https://doi.org/10.1016/0550-3213(85)90402-X},
reportNumber = "CERN-TH-3987/84",
doi = "10.1016/0550-3213(85)90402-X"
}

@article{Ladinsky:1993zn,
author = "Ladinsky, G. A. and Yuan, C. P.",
title = "{Nonperturbative regime in QCD resummation for gauge boson production at hadron colliders}",
eprint = "hep-ph/9311341",
archivePrefix = "arXiv",
reportNumber = "MSUHEP-93-20, MSUTH-93-20",
doi = "10.1103/PhysRevD.50.R4239",
journal = "Phys. Rev. D",
volume = "50",
pages = "R4239",
year = "1994"
}

@article{Ellis:1997sc,
author = "Ellis, R. Keith and Ross, D. A. and Veseli, Sinisa",
title = "{Vector boson production in hadronic collisions}",
eprint = "hep-ph/9704239",
archivePrefix = "arXiv",
reportNumber = "FERMILAB-PUB-97-082-T, SHEP-96-37",
doi = "10.1016/S0550-3213(97)00403-3",
journal = "Nucl. Phys. B",
volume = "503",
pages = "309--338",
year = "1997"
}

@article{Landry:1999an,
author = "Landry, F. and Brock, R. and Ladinsky, G. and Yuan, C. P.",
title = "{New fits for the non-perturbative parameters in the CSS resummation formalism}",
eprint = "hep-ph/9905391",
archivePrefix = "arXiv",
reportNumber = "CTEQ-904, MSUHEP-90414",
doi = "10.1103/PhysRevD.63.013004",
journal = "Phys. Rev. D",
volume = "63",
pages = "013004",
year = "2000"
}

@article{Qiu:2000hf,
 author = {Qiu, Jianwei and Zhang, Xiaofei},
title = "{Role of the nonperturbative input in QCD resummed Drell--Yan $Q_{T}$ distributions}",
eprint = "hep-ph/0012348",
archivePrefix = "arXiv",
doi = "10.1103/PhysRevD.63.114011",
journal = "Phys. Rev. D",
volume = "63",
pages = "114011",
year = "2001"
}

@article{Konychev:2005iy,
author = "Konychev, Anton V. and Nadolsky, Pavel M.",
title = "{Universality of the Collins-Soper-Sterman nonperturbative function in vector boson production}",
eprint = "hep-ph/0506225",
archivePrefix = "arXiv",
reportNumber = "ANL-HEP-PR-05-52",
doi = "10.1016/j.physletb.2005.12.063",
journal = "Phys. Lett. B",
volume = "633",
pages = "710--714",
year = "2006"
}

@article{Guzzi:2013aja,
author = "Guzzi, Marco and Nadolsky, Pavel M. and Wang, Bowen",
title = "{Nonperturbative contributions to a resummed leptonic angular distribution in inclusive neutral vector boson production}",
eprint = "1309.1393",
archivePrefix = "arXiv",
primaryClass = "hep-ph",
doi = "10.1103/PhysRevD.90.014030",
journal = "Phys. Rev. D",
volume = "90",
number = "1",
pages = "014030",
year = "2014"
}

@article{Collins:2014jpa,
    author = "Collins, John and Rogers, Ted",
    title = "{Understanding the large-distance behavior of transverse-momentum-dependent parton densities and the Collins-Soper evolution kernel}",
    eprint = "1412.3820",
    archivePrefix = "arXiv",
    primaryClass = "hep-ph",
    reportNumber = "YITP-WSB-14-45, JLAB-THY-15-2001",
    doi = "10.1103/PhysRevD.91.074020",
    journal = "Phys. Rev. D",
    volume = "91",
    number = "7",
    pages = "074020",
    year = "2015"
}

@article{Wei:2020glg,
    author = "Wei, Shuyi",
    title = "{Exploring the non-perturbative Sudakov factor via $Z^0$ -boson production in $pp$ collisions}",
    eprint = "2009.06514",
    archivePrefix = "arXiv",
    primaryClass = "hep-ph",
    doi = "10.1016/j.physletb.2021.136356",
    journal = "Phys. Lett. B",
    volume = "817",
    pages = "136356",
    year = "2021"
}

@article{Kulesza:2002rh,
    author = "Kulesza, Anna and Sterman, George F. and Vogelsang, Werner",
    title = "{Joint resummation in electroweak boson production}",
    eprint = "hep-ph/0202251",
    archivePrefix = "arXiv",
    reportNumber = "BNL-HET-01-43, BNL-NT-01-32, RBRC-232, YITP-SB-01-75",
    doi = "10.1103/PhysRevD.66.014011",
    journal = "Phys. Rev. D",
    volume = "66",
    pages = "014011",
    year = "2002"
}

@article{Caola:2021kzt,
    author = "Caola, Fabrizio and Ferrario Ravasio, Silvia and Limatola, Giovanni and Melnikov, Kirill and Nason, Paolo",
    title = "{On linear power corrections in certain collider observables}",
    eprint = "2108.08897",
    archivePrefix = "arXiv",
    primaryClass = "hep-ph",
    reportNumber = "OUTP-21-21P, TTP21-026, P3H-21-056",
    doi = "10.1007/JHEP01(2022)093",
    journal = "JHEP",
    volume = "01",
    pages = "093",
    year = "2022"
}

@article{FerrarioRavasio:2020guj,
    author = "Ferrario Ravasio, Silvia and Limatola, Giovanni and Nason, Paolo",
    title = "{Infrared renormalons in kinematic distributions for hadron collider processes}",
    eprint = "2011.14114",
    archivePrefix = "arXiv",
    primaryClass = "hep-ph",
    reportNumber = "OUTP-20-13P, IPPP/20/60",
    doi = "10.1007/JHEP06(2021)018",
    journal = "JHEP",
    volume = "06",
    pages = "018",
    year = "2021"
}

@article{Camarda:2019zyx,
    author = "Camarda, Stefano and others",
    title = "{DYTurbo: Fast predictions for Drell--Yan processes}",
    eprint = "1910.07049",
    archivePrefix = "arXiv",
    primaryClass = "hep-ph",
    doi = "10.1140/epjc/s10052-020-7757-5",
    journal = "Eur. Phys. J. C",
    volume = "80",
    number = "3",
    pages = "251",
    year = "2020",
    related        = "Camarda:2019zyx-err",
    relatedstring  = "Erratum:",
}

@article{Bozzi:2005wk,
    author = "Bozzi, Giuseppe and Catani, Stefano and de Florian, Daniel and Grazzini, Massimiliano",
    title = "{Transverse-momentum resummation and the spectrum of the Higgs boson at the LHC}",
    eprint = "hep-ph/0508068",
    archivePrefix = "arXiv",
    doi = "10.1016/j.nuclphysb.2005.12.022",
    journal = "Nucl. Phys. B",
    volume = "737",
    pages = "73--120",
    year = "2006"
}

@article{Catani:2015vma,
    author = "Catani, Stefano and de Florian, Daniel and Ferrera, Giancarlo and Grazzini, Massimiliano",
    title = "{Vector boson production at hadron colliders: transverse-momentum resummation and leptonic decay}",
    eprint = "1507.06937",
    archivePrefix = "arXiv",
    primaryClass = "hep-ph",
    reportNumber = "ICAS-02-15, TIF-UNIMI-2015-10, ZU-TH-24-15",
    doi = "10.1007/JHEP12(2015)047",
    journal = "JHEP",
    volume = "12",
    pages = "047",
    year = "2015"
}

@article{Collins:1981va,
    author = "Collins, John C. and Soper, Davison E.",
    title = {Back-to-back jets: Fourier transform from b to $k_T$},
    reportNumber = "OITS-153",
    doi = "10.1016/0550-3213(82)90453-9",
    journal = "Nucl. Phys. B",
    volume = "197",
    pages = "446--476",
    year = "1982"
}

@article{Camarda:2021jsw,
    author = "Camarda, Stefano and Cieri, Leandro and Ferrera, Giancarlo",
    title = "{Fiducial perturbative power corrections within the $\mathbf{q}_T$ subtraction formalism}",
    eprint = "2111.14509",
    archivePrefix = "arXiv",
    primaryClass = "hep-ph",
    reportNumber = "IFIC/21-51, FTUV-21-1129.5289",
    doi = "10.1140/epjc/s10052-022-10510-x",
    journal = "Eur. Phys. J. C",
    volume = "82",
    number = "6",
    pages = "575",
    year = "2022"
}

@article{Buckley:2014ana,
    author = {Buckley, Andy and Ferrando, James and Lloyd, Stephen and Nordstr\"om, Karl and Page, Ben and R\"ufenacht, Martin and Sch\"onherr, Marek and Watt, Graeme},
    title = "{LHAPDF6: parton density access in the LHC precision era}",
    eprint = "1412.7420",
    archivePrefix = "arXiv",
    primaryClass = "hep-ph",
    reportNumber = "GLAS-PPE-2014-05, MCNET-14-29, IPPP-14-111, DCPT-14-222",
    doi = "10.1140/epjc/s10052-015-3318-8",
    journal = "Eur. Phys. J. C",
    volume = "75",
    pages = "132",
    year = "2015"
}

@article{Boughezal:2015ded,
    author = "Boughezal, Radja and Campbell, John M. and Ellis, R. Keith and Focke, Christfried and Giele, Walter T. and Liu, Xiaohui and Petriello, Frank",
    title = "{Z-boson Production in Association with a Jet at Next-to-Next-To-Leading Order in Perturbative QCD}",
    eprint = "1512.01291",
    archivePrefix = "arXiv",
    primaryClass = "hep-ph",
    reportNumber = "FERMILAB-PUB-15-519-T, IPPP-15-79",
    doi = "10.1103/PhysRevLett.116.152001",
    journal = "Phys. Rev. Lett.",
    volume = "116",
    number = "15",
    pages = "152001",
    year = "2016"
}

@article{Alekhin:2014irh,
    author = "Alekhin, S. and others",
    title = "{HERAFitter}",
    eprint = "1410.4412",
    archivePrefix = "arXiv",
    primaryClass = "hep-ph",
    reportNumber = "DESY-14-188, DESY-REPORT-14-188, FERMILAB-PUB-14-603-CMS",
    doi = "10.1140/epjc/s10052-015-3480-z",
    journal = "Eur. Phys. J. C",
    volume = "75",
    number = "7",
    pages = "304",
    year = "2015"
}

@article{HERAFitterdevelopersTeam:2015cre,
    author = "Camarda, S. and others",
    collaboration = "HERAFitter developers' Team",
    title = "{QCD analysis of $W$- and $Z$-boson production at Tevatron}",
    eprint = "1503.05221",
    archivePrefix = "arXiv",
    primaryClass = "hep-ph",
    reportNumber = "DESY-15-035, FERMILAB-PUB-15-209-PPD, DESY-REPORT-15-035, DESY Report 15-035",
    doi = "10.1140/epjc/s10052-015-3655-7",
    journal = "Eur. Phys. J. C",
    volume = "75",
    number = "9",
    pages = "458",
    year = "2015"
}

@article{Sjostrand:2014zea,
    author = {Sj\"ostrand, Torbj\"orn and Ask, Stefan and Christiansen, Jesper R. and Corke, Richard and Desai, Nishita and Ilten, Philip and Mrenna, Stephen and Prestel, Stefan and Rasmussen, Christine O. and Skands, Peter Z.},
    title = "{An introduction to PYTHIA 8.2}",
    eprint = "1410.3012",
    archivePrefix = "arXiv",
    primaryClass = "hep-ph",
    reportNumber = "LU-TP-14-36, MCNET-14-22, CERN-PH-TH-2014-190, FERMILAB-PUB-14-316-CD, DESY-14-178, SLAC-PUB-16122",
    doi = "10.1016/j.cpc.2015.01.024",
    journal = "Comput. Phys. Commun.",
    volume = "191",
    pages = "159--177",
    year = "2015"
}

@article{Billis:2019evv,
    author = "Billis, Georgios and Tackmann, Frank J. and Talbert, Jim",
    title = "{Higher-Order Sudakov resummation in coupled gauge theories}",
    eprint = "1907.02971",
    archivePrefix = "arXiv",
    primaryClass = "hep-ph",
    reportNumber = "DESY-19-128, DESY 19-128",
    doi = "10.1007/JHEP03(2020)182",
    journal = "JHEP",
    volume = "03",
    pages = "182",
    year = "2020"
}

@article{ParticleDataGroup:1986kuw,
    author = "Aguilar-Benitez, M. and others",
    collaboration = "Particle Data Group",
    title = "{Review of Particle Properties. Particle Data Group}",
    journal = "Phys. Lett. B",
    volume = "170",
    pages = "1--350",
    year = "1986"
}

@article{ParticleDataGroup:2022pth,
    author = "Workman, R. L. and others",
    collaboration = "Particle Data Group",
    title = "{Review of Particle Physics}",
    doi = "10.1093/ptep/ptac097",
    journal = "PTEP",
    volume = "2022",
    pages = "083C01",
    year = "2022"
}

@article{Gross:1973id,
    author = "Gross, David J. and Wilczek, Frank",
    title = {Ultraviolet Behavior of Non-Abelian Gauge Theories},
    doi = "10.1103/PhysRevLett.30.1343",
    journal = "Phys. Rev. Lett.",
    volume = "30",
    pages = "1343--1346",
    year = "1973"
}

@article{Politzer:1973fx,
    author = "Politzer, H. David",
    title = "{Reliable Perturbative Results for Strong Interactions?}",
    doi = "10.1103/PhysRevLett.30.1346",
    journal = "Phys. Rev. Lett.",
    volume = "30",
    pages = "1346--1349",
    year = {1973}
}

@article{Heinemeyer:2021rgq,
    author = {Heinemeyer, Sven and Jadach, Stanislaw and Reuter, J\"urgen},
    title = "{Theory requirements for SM Higgs and EW precision physics at the FCC-ee}",
    eprint = "2106.11802",
    archivePrefix = "arXiv",
    primaryClass = "hep-ph",
    reportNumber = "DESY-21-100",
    doi = "10.1140/epjp/s13360-021-01875-1",
    journal = "Eur. Phys. J. Plus",
    volume = "136",
    number = "9",
    pages = "911",
    year = "2021"
}

@article{dEnterria:2020cpv,
    author = "d'Enterria, David and Jacobsen, Villads",
    title = "{Improved strong coupling determinations from hadronic decays of electroweak bosons at N$^3$LO accuracy}",
    eprint = "2005.04545",
    archivePrefix = "arXiv",
    primaryClass = "hep-ph",
    month = "5",
    year = "2020"
}

@article{Degrassi:2012ry,
    author = "Degrassi, Giuseppe and Di Vita, Stefano and Elias-Miro, Joan and Espinosa, Jose R. and Giudice, Gian F. and Isidori, Gino and Strumia, Alessandro",
    title = "{Higgs mass and vacuum stability in the Standard Model at NNLO}",
    eprint = "1205.6497",
    archivePrefix = "arXiv",
    primaryClass = "hep-ph",
    reportNumber = "CERN-PH-TH-2012-134, RM3-TH-12-9",
    doi = "10.1007/JHEP08(2012)098",
    journal = "JHEP",
    volume = "08",
    pages = "098",
    year = "2012"
}

@article{Pich:2016bdg,
    author = "Pich, Antonio and Rodr\'iguez-S\'anchez, Antonio",
    title = "{Determination of the QCD coupling from ALEPH $\tau$ decay data}",
    eprint = "1605.06830",
    archivePrefix = "arXiv",
    primaryClass = "hep-ph",
    reportNumber = "IFIC-16-13, FTUV-16-0522",
    doi = "10.1103/PhysRevD.94.034027",
    journal = "Phys. Rev. D",
    volume = "94",
    number = "3",
    pages = "034027",
    year = "2016"
}

@article{Davier:2013sfa,
    author = {Davier, Michel and H\"ocker, Andreas and Malaescu, Bogdan and Yuan, Chang-Zheng and Zhang, Zhiqing},
    title = "{Update of the ALEPH non-strange spectral functions from hadronic $\tau$ decays}",
    eprint = "1312.1501",
    archivePrefix = "arXiv",
    primaryClass = "hep-ex",
    reportNumber = "LAL-13-390",
    doi = "10.1140/epjc/s10052-014-2803-9",
    journal = "Eur. Phys. J. C",
    volume = "74",
    number = "3",
    pages = "2803",
    year = "2014"
}

@article{Boito:2018yvl,
    author = "Boito, Diogo and Golterman, Maarten and Keshavarzi, Alexander and Maltman, Kim and Nomura, Daisuke and Peris, Santiago and Teubner, Thomas",
    title = "{Strong coupling from $e^+e^-\to$ hadrons below charm}",
    eprint = "1805.08176",
    archivePrefix = "arXiv",
    primaryClass = "hep-ph",
    reportNumber = "KEK-TH-2049, LTH-1156",
    doi = "10.1103/PhysRevD.98.074030",
    journal = "Phys. Rev. D",
    volume = "98",
    number = "7",
    pages = "074030",
    year = "2018"
}

@article{Boito:2020xli,
    author = "Boito, Diogo and Golterman, Maarten and Maltman, Kim and Peris, Santiago and Rodrigues, Marcus V. and Schaaf, Wilder",
    title = "{Strong coupling from an improved $\tau$ vector isovector spectral function}",
    eprint = "2012.10440",
    archivePrefix = "arXiv",
    primaryClass = "hep-ph",
    doi = "10.1103/PhysRevD.103.034028",
    journal = "Phys. Rev. D",
    volume = "103",
    number = "3",
    pages = "034028",
    year = "2021"
}

@article{Camarda:2022qdg,
    author = "Camarda, Stefano and Ferrera, Giancarlo and Schott, Matthias",
    title = "{Determination of the strong-coupling constant from the $Z$-boson transverse-momentum distribution}",
    eprint = "2203.05394",
    archivePrefix = "arXiv",
    primaryClass = "hep-ph",
    month = "3",
    year = "2022"
}

@article{Tafat:2001in,
    author = "Tafat, Sofiane",
    title = "{Nonperturbative corrections to the Drell--Yan transverse momentum distribution}",
    eprint = "hep-ph/0102237",
    archivePrefix = "arXiv",
    reportNumber = "LPT-ORSAY-01-12",
    doi = "10.1088/1126-6708/2001/05/004",
    journal = "JHEP",
    volume = "05",
    pages = "004",
    year = "2001"
}

@article{Collins:1977iv,
    author = "Collins, John C. and Soper, Davison E.",
    title = "{Angular Distribution of Dileptons in High-Energy Hadron Collisions}",
    reportNumber = "Print-77-0288 (PRINCETON)",
    doi = "10.1103/PhysRevD.16.2219",
    journal = "Phys. Rev. D",
    volume = "16",
    pages = "2219",
    year = "1977"
}

@article{Mirkes:1992hu,
    author = "Mirkes, E.",
    title = "{Angular decay distribution of leptons from W-bosons at NLO in hadronic collisions}",
    reportNumber = "TTP-92-12",
    doi = "10.1016/0550-3213(92)90046-E",
    journal = "Nucl. Phys. B",
    volume = "387",
    pages = "3--85",
    year = "1992"
}

@article{Herzog:2018kwj,
    author = "Herzog, F. and Moch, S. and Ruijl, B. and Ueda, T. and Vermaseren, J. A. M. and Vogt, A.",
    title = "{Five-loop contributions to low-N non-singlet anomalous dimensions in QCD}",
    eprint = "1812.11818",
    archivePrefix = "arXiv",
    primaryClass = "hep-ph",
    reportNumber = "Nikhef 2018-066, DESY 18-227, DESY-18-227, LTH 1191",
    doi = "10.1016/j.physletb.2019.01.060",
    journal = "Phys. Lett. B",
    volume = "790",
    pages = "436--443",
    year = "2019"
}

@article{Duhr:2022yyp,
    author = "Duhr, Claude and Mistlberger, Bernhard and Vita, Gherardo",
    title = "{Four-Loop Rapidity Anomalous Dimension and Event Shapes to Fourth Logarithmic Order}",
    eprint = "2205.02242",
    archivePrefix = "arXiv",
    primaryClass = "hep-ph",
    reportNumber = "BONN-TH-2022-11, SLAC-PUB-17675",
    doi = "10.1103/PhysRevLett.129.162001",
    journal = "Phys. Rev. Lett.",
    volume = "129",
    number = "16",
    pages = "162001",
    year = "2022"
}

@article{Moult:2022xzt,
    author = "Moult, Ian and Zhu, Hua Xing and Zhu, Yu Jiao",
    title = "{The four loop QCD rapidity anomalous dimension}",
    eprint = "2205.02249",
    archivePrefix = "arXiv",
    primaryClass = "hep-ph",
    doi = "10.1007/JHEP08(2022)280",
    journal = "JHEP",
    volume = "08",
    pages = "280",
    year = "2022"
}

@article{Bondarenko:2022mbi,
    author = "Bondarenko, Serge and Dydyshka, Yahor and Kalinovskaya, Lidia and Sadykov, Renat and Yermolchyk, Vitaly",
    title = "{Hadron-hadron collision mode in ReneSANCe-v1.3.0}",
    eprint = "2207.04332",
    archivePrefix = "arXiv",
    primaryClass = "hep-ph",
    doi = "10.1016/j.cpc.2022.108646",
    journal = "Comput. Phys. Commun.",
    volume = "285",
    pages = "108646",
    year = "2023"
}

@article{n4lla,
    author = "Camarda, Stefano and Cieri, Leandro and Ferrera, Giancarlo",
    title = {Drell--Yan lepton-pair production: $q_T$ resummation at N$^4$LL accuracy},
    eprint = "2303.12781",
    archivePrefix = "arXiv",
    primaryClass = "hep-ph",
    journal = {Physics Letters B},
    volume = {845},
    pages = {138125},
    year = {2023},
    doi = {https://doi.org/10.1016/j.physletb.2023.138125},
}

@article{McGowan:2022nag,
    author = "McGowan, J. and Cridge, T. and Harland-Lang, L. A. and Thorne, R. S.",
    title = "{Approximate N$^{3}$LO parton distribution functions with theoretical uncertainties: MSHT20aN$^3$LO PDFs}",
    eprint = "2207.04739",
    archivePrefix = "arXiv",
    primaryClass = "hep-ph",
    doi = "10.1140/epjc/s10052-023-11236-0",
    journal = "Eur. Phys. J. C",
    volume = "83",
    number = "3",
    pages = "185",
    year = "2023"
}

@article{Bertone:2022sso,
    author = "Bertone, V. and Bozzi, G. and Hautmann, F.",
    title = "{Perturbative hysteresis and emergent resummation scales}",
    eprint = "2202.03380",
    archivePrefix = "arXiv",
    primaryClass = "hep-ph",
    doi = "10.1103/PhysRevD.105.096003",
    journal = "Phys. Rev. D",
    volume = "105",
    number = "9",
    pages = "096003",
    year = "2022"
}

@article{Baikov:2016tgj,
    author = {Baikov, P. A. and Chetyrkin, K. G. and K\"uhn, J. H.},
    title = "{Five-Loop Running of the QCD coupling constant}",
    eprint = "1606.08659",
    archivePrefix = "arXiv",
    primaryClass = "hep-ph",
    reportNumber = "TTP16-026",
    doi = "10.1103/PhysRevLett.118.082002",
    journal = "Phys. Rev. Lett.",
    volume = "118",
    number = "8",
    pages = "082002",
    year = "2017"
}

@article{Chakraborty:2022yan,
    author = "Chakraborty, Amlan and Huber, Tobias and Lee, Roman N. and von Manteuffel, Andreas and Schabinger, Robert M. and Smirnov, Alexander V. and Smirnov, Vladimir A. and Steinhauser, Matthias",
    title = "{Hbb vertex at four loops and hard matching coefficients in SCET for various currents}",
    eprint = "2204.02422",
    archivePrefix = "arXiv",
    primaryClass = "hep-ph",
    reportNumber = "MSUHEP-22-014, P3H-22-034, TTP22-020, SI-HEP-2022-06",
    doi = "10.1103/PhysRevD.106.074009",
    journal = "Phys. Rev. D",
    volume = "106",
    number = "7",
    pages = "074009",
    year = "2022"
}

@article{Cieri:2018sfk,
    author = "Cieri, Leandro and Ferrera, Giancarlo and Sborlini, German F. R.",
    title = "{Combining QED and QCD transverse-momentum resummation for Z boson production at hadron colliders}",
    eprint = "1805.11948",
    archivePrefix = "arXiv",
    primaryClass = "hep-ph",
    reportNumber = "TIF-UNIMI-2018-4",
    doi = "10.1007/JHEP08(2018)165",
    journal = "JHEP",
    volume = "08",
    pages = "165",
    year = "2018"
}

@article{Bailey:2020ooq,
    author = "Bailey, S. and Cridge, T. and Harland-Lang, L. A. and Martin, A. D. and Thorne, R. S.",
    title = "{Parton distributions from LHC, HERA, Tevatron and fixed target data: MSHT20 PDFs}",
    eprint = "2012.04684",
    archivePrefix = "arXiv",
    primaryClass = "hep-ph",
    reportNumber = "IPPP/20/58",
    doi = "10.1140/epjc/s10052-021-09057-0",
    journal = "Eur. Phys. J. C",
    volume = "81",
    number = "4",
    pages = "341",
    year = "2021"
}

@article{NNPDF:2021njg,
    author = "Ball, Richard D. and others",
    collaboration = "NNPDF",
    title = "{The path to proton structure at 1\% accuracy}",
    eprint = "2109.02653",
    archivePrefix = "arXiv",
    primaryClass = "hep-ph",
    reportNumber = "Edinburgh 2021/12, Nikhef-2021-013, TIF-UNIMI-2021-11",
    doi = "10.1140/epjc/s10052-022-10328-7",
    journal = "Eur. Phys. J. C",
    volume = "82",
    number = "5",
    pages = "428",
    year = "2022"
}

@article{H1:2015ubc,
    author = "Abramowicz, H. and others",
    collaboration = "H1, ZEUS",
    title = "{Combination of measurements of inclusive deep inelastic ${e^{\pm }p}$ scattering cross sections and QCD analysis of HERA data}",
    eprint = "1506.06042",
    archivePrefix = "arXiv",
    primaryClass = "hep-ex",
    reportNumber = "DESY-15-039",
    doi = "10.1140/epjc/s10052-015-3710-4",
    journal = "Eur. Phys. J. C",
    volume = "75",
    number = "12",
    pages = "580",
    year = "2015"
}

@article{Hou:2019efy,
    author = "Hou, Tie-Jiun and others",
    title = "{New CTEQ global analysis of quantum chromodynamics with high-precision data from the LHC}",
    eprint = "1912.10053",
    archivePrefix = "arXiv",
    primaryClass = "hep-ph",
    reportNumber = "MSUHEP-19-025, PITT-PACC-1911, SMU-HEP-19-03",
    doi = "10.1103/PhysRevD.103.014013",
    journal = "Phys. Rev. D",
    volume = "103",
    number = "1",
    pages = "014013",
    year = "2021"
}

@software{Agarwal_Ceres_Solver_2022,
  author = {Agarwal, Sameer and Mierle, Keir and The Ceres Solver Team},
  title = {{Ceres Solver}},
  license = {Apache-2.0},
  url = {https://github.com/ceres-solver/ceres-solver},
  version = {2.1},
  year = {2022},
  month = {3}
}

@article{Weinberg:1973xwm,
    author = "Weinberg, Steven",
    title = "{New Approach to the Renormalization Group}",
    doi = "10.1103/PhysRevD.8.3497",
    journal = "Phys. Rev. D",
    volume = "8",
    pages = "3497--3509",
    year = "1973"
}

@article{Thorne:2012az,
    author = "Thorne, R. S.",
    title = "{Effect of changes of variable flavor number scheme on parton distribution functions and predicted cross sections}",
    eprint = "1201.6180",
    archivePrefix = "arXiv",
    primaryClass = "hep-ph",
    doi = "10.1103/PhysRevD.86.074017",
    journal = "Phys. Rev. D",
    volume = "86",
    pages = "074017",
    year = "2012"
}

@article{Thorne:2006qt,
    author = "Thorne, R. S.",
    title = {Variable-flavor number scheme for next-to-next-to-leading order},
    eprint = "hep-ph/0601245",
    archivePrefix = "arXiv",
    reportNumber = "CAVENDISH-HEP-2006-01",
    doi = "10.1103/PhysRevD.73.054019",
    journal = "Phys. Rev. D",
    volume = "73",
    pages = "054019",
    year = "2006"
}

@article{Thorne:1997ga,
    author = "Thorne, R. S. and Roberts, R. G.",
    title = {Ordered analysis of heavy flavor production in deep-inelastic scattering},
    eprint = "hep-ph/9709442",
    archivePrefix = "arXiv",
    reportNumber = "RAL-TR-97-049",
    doi = "10.1103/PhysRevD.57.6871",
    journal = "Phys. Rev. D",
    volume = "57",
    pages = "6871--6898",
    year = "1998"
}

@article{Bonvini:2019wxf,
    author = "Bonvini, Marco and Giuli, Francesco",
    title = "{A new simple PDF parametrization: Improved description of the HERA data}",
    eprint = "1902.11125",
    archivePrefix = "arXiv",
    primaryClass = "hep-ph",
    doi = "10.1140/epjp/i2019-12872-x",
    journal = "Eur. Phys. J. Plus",
    volume = "134",
    number = "10",
    pages = "531",
    year = "2019"
}

@article{Paukkunen:2014zia,
    author = "Paukkunen, Hannu and Zurita, Pia",
    title = "{PDF reweighting in the Hessian matrix approach}",
    eprint = "1402.6623",
    archivePrefix = "arXiv",
    primaryClass = "hep-ph",
    doi = "10.1007/JHEP12(2014)100",
    journal = "JHEP",
    volume = "12",
    pages = "100",
    year = "2014"
}

@article{Forte:2020pyp,
    author = "Forte, Stefano and Kassabov, Zahari",
    title = "{Why $\alpha _s$ cannot be determined from hadronic processes without simultaneously determining the parton distributions}",
    eprint = "2001.04986",
    archivePrefix = "arXiv",
    primaryClass = "hep-ph",
    reportNumber = "TIF-UNIMI-2020-2, CAVENDISH-HEP-20-01",
    doi = "10.1140/epjc/s10052-020-7748-6",
    journal = "Eur. Phys. J. C",
    volume = "80",
    number = "3",
    pages = "182",
    year = "2020"
}

@article{Schweitzer:2012hh,
    author = "Schweitzer, P. and Strikman, M. and Weiss, C.",
    title = "{Intrinsic transverse momentum and parton correlations from dynamical chiral symmetry breaking}",
    eprint = "1210.1267",
    archivePrefix = "arXiv",
    primaryClass = "hep-ph",
    reportNumber = "JLAB-THY-12-1641",
    doi = "10.1007/JHEP01(2013)163",
    journal = "JHEP",
    volume = "01",
    pages = "163",
    year = "2013"
}

@article{Catani:1990rr,
    author = "Catani, S. and Webber, B. R. and Marchesini, G.",
    title = "{QCD coherent branching and semi-inclusive processes at large x}",
    reportNumber = "CAVENDISH-HEP-90-11, UPRF-90-280",
    doi = "10.1016/0550-3213(91)90390-J",
    journal = "Nucl. Phys. B",
    volume = "349",
    pages = "635--654",
    year = "1991"
}

@article{Herren:2017osy,
    author = "Herren, Florian and Steinhauser, Matthias",
    title = "{Version 3 of RunDec and CRunDec}",
    eprint = "1703.03751",
    archivePrefix = "arXiv",
    primaryClass = "hep-ph",
    reportNumber = "TTP17-011",
    doi = "10.1016/j.cpc.2017.11.014",
    journal = "Comput. Phys. Commun.",
    volume = "224",
    pages = "333--345",
    year = "2018"
}

@article{H1:2017bml,
    author = "Andreev, V. and others",
    collaboration = "H1",
    title = "{Determination of the strong coupling constant $\alpha_s(m_Z)$ in next-to-next-to-leading order QCD using H1 jet cross section measurements}",
    eprint = "1709.07251",
    archivePrefix = "arXiv",
    primaryClass = "hep-ex",
    reportNumber = "DESY-17-137, DESY17-137",
    doi = "10.1140/epjc/s10052-017-5314-7",
    journal = "Eur. Phys. J. C",
    volume = "77",
    number = "11",
    pages = "791",
    year = "2017",
    related        = "H1:2017bml-err",
    relatedstring  = "Erratum:",
}

@article{Britzger:2019kkb,
    author = "Britzger, D. and others",
    title = "{Calculations for deep inelastic scattering using fast interpolation grid techniques at NNLO in QCD and the extraction of $\alpha_s$ from HERA data}",
    eprint = "1906.05303",
    archivePrefix = "arXiv",
    primaryClass = "hep-ph",
    reportNumber = "CERN-TH-2019-079, CFTP/19-020, IPPP/19/44, MPP-2019-114, ZU-TH 29/19",
    doi = "10.1140/epjc/s10052-019-7351-x",
    journal = "Eur. Phys. J. C",
    volume = "79",
    number = "10",
    pages = "845",
    year = "2019",
    related        = "Britzger:2019kkb-err",
    relatedstring  = "Erratum:",
}

@article{Catani:1989ne,
    author = "Catani, S. and Trentadue, L.",
    title = "{Resummation of the QCD Perturbative Series for Hard Processes}",
    reportNumber = "DFF-93/3/89",
    doi = "10.1016/0550-3213(89)90273-3",
    journal = "Nucl. Phys. B",
    volume = "327",
    pages = "323--352",
    year = "1989"
}

@article{Cridge:2021qfd,
    author = "Cridge, T. and Harland-Lang, L. A. and Martin, A. D. and Thorne, R. S.",
    title = "{An investigation of the $\alpha _S$ and heavy quark mass dependence in the MSHT20 global PDF analysis}",
    eprint = "2106.10289",
    archivePrefix = "arXiv",
    primaryClass = "hep-ph",
    doi = "10.1140/epjc/s10052-021-09533-7",
    journal = "Eur. Phys. J. C",
    volume = "81",
    number = "8",
    pages = "744",
    year = "2021"
}

@article{Vladimirov:2020umg,
    author = "Vladimirov, Alexey A.",
    title = "{Self-Contained Definition of the Collins-Soper Kernel}",
    eprint = "2003.02288",
    archivePrefix = "arXiv",
    primaryClass = "hep-ph",
    doi = "10.1103/PhysRevLett.125.192002",
    journal = "Phys. Rev. Lett.",
    volume = "125",
    number = "19",
    pages = "192002",
    year = "2020"
}


%% file: ATLAS.bib
@Article{PERF-2007-01,
    author         = "{ATLAS Collaboration}",
    title          = "{The ATLAS Experiment at the CERN Large Hadron Collider}",
    journal        = "JINST",
    volume         = "3",
    year           = "2008",
    pages          = "S08003",
    doi            = "10.1088/1748-0221/3/08/S08003",
    primaryClass   = "hep-ex",
}

@Article{STDM-2012-23,
    author         = "{ATLAS Collaboration}",
    title          = "{Measurement of the \(Z/\gamma^*\) boson transverse momentum distribution in \(pp\) collisions at \(\sqrt{s} = 7\,\text{TeV}\) with the ATLAS detector}",
    journal        = "JHEP",
    volume         = "09",
    year           = "2014",
    pages          = "145",
    doi            = "10.1007/JHEP09(2014)145",
    reportNumber   = "CERN-PH-EP-2014-075",
    eprint         = "1406.3660",
    archivePrefix  = "arXiv",
    primaryClass   = "hep-ex",
}

@Article{DAPR-2013-01,
    author         = "{ATLAS Collaboration}",
    title          = "{Luminosity determination in \(pp\) collisions at \(\sqrt{s} = 8\,\text{TeV}\) using the ATLAS detector at the LHC}",
    journal        = "Eur. Phys. J. C",
    volume         = "76",
    year           = "2016",
    pages          = "653",
    doi            = "10.1140/epjc/s10052-016-4466-1",
    eprint         = "1608.03953",
    archivePrefix  = "arXiv",
    primaryClass   = "hep-ex",
}

@Article{STDM-2014-12,
    author         = "{ATLAS Collaboration}",
    title          = "{Measurement of the transverse momentum and \(\phi^*_{\eta}\) distributions of Drell-Yan lepton pairs in proton--proton collisions at \(\sqrt{s} = 8\,\text{TeV}\) with the ATLAS detector}",
    journal        = "Eur. Phys. J. C",
    volume         = "76",
    year           = "2016",
    pages          = "291",
    doi            = "10.1140/epjc/s10052-016-4070-4",
    reportNumber   = "CERN-PH-EP-2015-275",
    eprint         = "1512.02192",
    archivePrefix  = "arXiv",
    primaryClass   = "hep-ex",
}

@Article{STDM-2018-51,
    author         = "{ATLAS Collaboration}",
    title          = "{Determination of the strong coupling constant from transverse energy--energy correlations in multijet events at \(\sqrt{s} = 13\,\text{TeV}\) with the ATLAS detector}",
    journal        = "JHEP",
    volume         = "07",
    year           = "2023",
    pages          = "085",
    doi            = "10.1007/JHEP07(2023)085",
    reportNumber   = "CERN-EP-2022-282",
    eprint         = "2301.09351",
    archivePrefix  = "arXiv",
    primaryClass   = "hep-ex",
}

@Booklet{ATL-SOFT-PUB-2021-001,
    author         = "{ATLAS Collaboration}",
    title          = "{The ATLAS Collaboration Software and Firmware}",
    howpublished   = "{ATL-SOFT-PUB-2021-001}",
    url            = "https://cds.cern.ch/record/2767187",
    year           = "2021",
}

@Booklet{ATL-SOFT-PUB-2023-001,
    author         = "{ATLAS Collaboration}",
    title          = "{ATLAS Computing Acknowledgements}",
    howpublished   = "{ATL-SOFT-PUB-2023-001}",
    url            = "https://cds.cern.ch/record/2869272",
    year           = "2023",
}


%% file: CMS.bib
@Article{CMS-SMP-13-013,
    author         = "{CMS Collaboration}",
    title          = "{Measurement of the \(Z\) boson differential cross section in transverse momentum and rapidity in proton--proton collisions at \(8\,\text{TeV}\)}",
    journal        = "Phys. Lett. B",
    volume         = "749",
    year           = "2015",
    pages          = "187",
    doi            = "10.1016/j.physletb.2015.07.065",
    reportNumber   = "CERN-PH-EP-2015-059",
    eprint         = "1504.03511",
    archivePrefix  = "arXiv",
    primaryClass   = "hep-ex",
}

@Article{CMS-TOP-17-001,
    author         = "{CMS Collaboration}",
    title          = "{Measurement of the \(t\bar{t}\) production cross section, the top quark mass, and the strong coupling constant using dilepton events in \(pp\) collisions at \(\sqrt{s} = 13\,\text{TeV}\)}",
    journal        = "Eur. Phys. J. C",
    volume         = "79",
    year           = "2019",
    pages          = "368",
    doi            = "10.1140/epjc/s10052-019-6863-8",
    reportNumber   = "CERN-EP-2018-317",
    eprint         = "1812.10505",
    archivePrefix  = "arXiv",
    primaryClass   = "hep-ex",
}

@Article{CMS-SMP-20-011,
    author         = "{CMS Collaboration}",
    title          = "{Measurement and QCD analysis of double-differential inclusive jet cross sections in proton--proton collisions at \(\sqrt{s} = 13\,\text{TeV}\)}",
    journal        = "JHEP",
    volume         = "02",
    year           = "2021",
    pages          = "142",
    doi            = "10.1007/JHEP02(2022)142",
    reportNumber   = "CERN-EP-2021-221",
    eprint         = "2111.10431",
    archivePrefix  = "arXiv",
    primaryClass   = "hep-ex",
}
